\def\preprint{1}                %  preprint
\def\comment#1{}
\if\preprint1
        \documentclass[useAMS,usenatbib]{mn2e}
        \usepackage{natbib}
        \usepackage{times}
        \usepackage{graphicx}
         \usepackage{url}
        \usepackage{calc}
        \usepackage{amssymb,amsmath}
        \usepackage{rotating}
         \usepackage{lscape}
         \bibpunct{(}{)}{;}{a}{}{,} % to follow the A&A style
         \usepackage{longtable}
\else
        \documentstyle[astrop-bib,referee,times]{mn2e}
        \newcommand{\includegraphics}[1]{}
        
\fi

\def\oversim#1#2{\lower0.5pt\vbox{\baselineskip0pt \lineskip-0.5pt
     \ialign{$\mathsurround0pt #1\hfil##\hfil$\crcr#2\crcr\sim\crcr}}}
\title[Submm polarisation of CRL 618 and OH 231.8+4.2]{Submillimeter polarisation and magnetic field properties in the envelopes of proto-planetary nebulae CRL 618 and OH 231.8+4.2}
\author[L. Sabin et al.]{L. Sabin$^{1}$\thanks{E-mail:lsabin@astro.iam.udg.mx (LS)}, Q. Zhang$^{2}$, A.~A. Zijlstra$^{3}$, N.A. Patel$^{2}$, R. V\'{a}zquez$^{4}$, B.~A. Zauderer$^{2}$, 
\newauthor  M.~E. Contreras$^{4}$ and P.~F. Guill\'{e}n$^{4}$\\ 
$^{1}${Instituto de Astonom{\'i}a y Meteorolog{\'i}a, Departamento de F{\'i}sica, CUCEI, Universidad de Guadalajara, Av. Vallarta 2602, C.P. 44130, Guadalajara, Jal., Mexico}\\
$^{2}$Harvard-Smithsonian Center for Astrophysics, 60 Garden Street, Cambridge, MA 02138, USA\\
$^{3}$Jodrell Bank Centre for Astrophysics, Alan Turing Building, University of Manchester, Manchester, M13 9PL, UK\\
$^{4}$Instituto de Astronom\'{i}a, Universidad Nacional Aut\'{o}noma de M\'{e}xico, Apdo. Postal 877, 22800 Ensenada, B. C, Mexico.\\
}
\begin{document}

\date{Accepted 2013 November 29. Received 2013 November 21; in original form 2013 September 4}

\pagerange{\pageref{firstpage}--\pageref{lastpage}} \pubyear{2013}

\maketitle

\label{firstpage}

\begin{abstract}
We have carried out continuum and line polarisation observations
of two Proto-planetary nebulae (PPNe), CRL~618 and OH~231.8+4.2,
using the Submillimeter Array (SMA) in its compact configuration. The 
frequency range of observations, 330--345 GHz, includes the CO($J$=3$\rightarrow$2) line emission.
CRL 618 and OH~231.8+4.2 show quadrupolar and bipolar optical lobes, 
respectively, surrounded by a dusty envelope reminiscent of their 
AGB phase. We report a detection of dust continuum polarised emission in both PPNe above $4\sigma$ but no molecular line polarisation detection above a $3\sigma$ limit. OH~231.8+4.2 is slightly more polarised on average than CRL~618 with a mean
fractional polarisation of 4.3 and 0.3 per cent, respectively. This agrees with the
previous finding that silicate dust shows higher polarisation than carbonaceous
dust. In both objects, an anti-correlation between the fractional polarisation and
the intensity is observed. Neither PPNe show a well defined toroidal
equatorial field, rather the field is generally well aligned and organised along
the polar direction. This is clearly seen in CRL 618 while in the case of
OH~231.8+4.2, the geometry indicates an X-shaped structure
coinciding overall with a dipole/polar configuration. However in the later case, the presence of a fragmented and weak toroidal field should not be discarded. Finally, in both PPNe, we observed that the well organised magnetic field is parallel with the major axis of the $^{12}$CO outflow. This alignment could indicate the presence of a magnetic outflow launching mechanism. Based on our new high resolution data we propose two scenarios to explain the evolution of the magnetic field in evolved stars.
\end{abstract}


\section{Introduction}

\begin{figure}
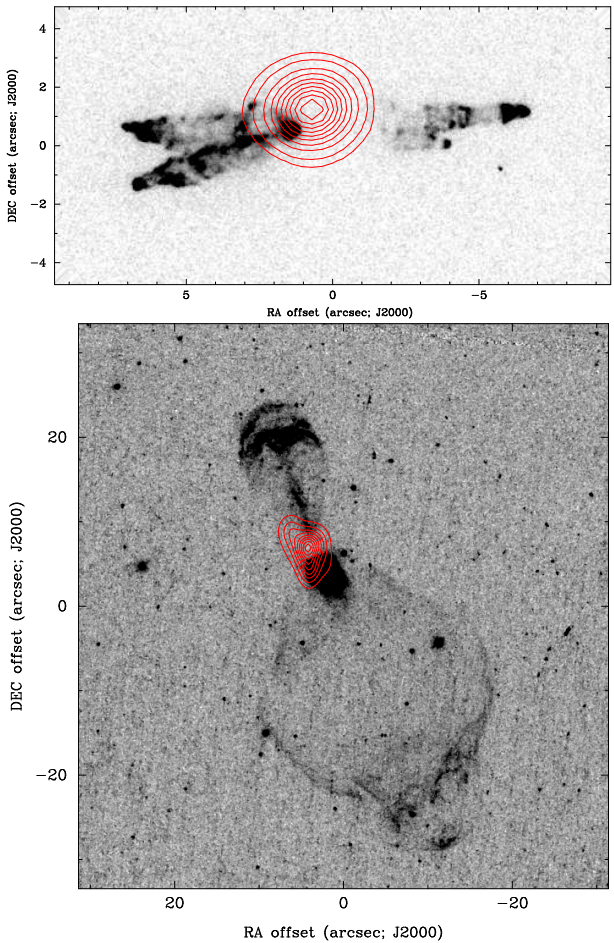

\begin{center}
%\vspace{1cm}
{\includegraphics[height=8.05cm,angle=-90]{CRL618.ps}}
{\includegraphics[height=8cm,angle=-90]{OH231.ps}}
\caption{\label{HST} H$\alpha$ images of CRL 618 (top) and OH 231.8+4.2 (bottom) from the Hubble Space Telescope with their respective dust continuum emission [(red contours in steps of $\sim$3$\sigma$ $\times$\,(0.5,1,2,3,4,5,6,7,8,9)] from the SMA observations, superimposed. Both PPNe show a clear asymmetry and the submm emission is roughly elongated in the direction of the outflows ({\it see more details in the text}). The positional offset of the continuum peak and (0,0) position is stated  in $\S$2 and $\S$3 for both objects. North is up and East is left in all frames. }
\end{center}
\end{figure}

Proto-planetary nebulae (PPNe) are the by-products of low and intermediate
mass stars ($\sim$0.8--8 M$_{\odot}$) in transition between the asymptotic giant
branch (AGB) and the planetary nebula (PN) phases. At this stage of stellar
evolution the central star is not hot enough to fully ionise the still present
circumstellar envelope. PPNe therefore have massive molecular and dust shells,
relics of past AGB mass loss events \citep{Kwok1993}. Consequently
observational studies based on the thermal emission of the dust continuum
and molecular species have been conducted over the years at submillimeter,
millimeter and centimetre wavelengths \citep[e.g.][]{Sanchez1998,Woods2005,Bujarrabal2012}. These studies revealed the properties of the dust grains and molecules, the distribution and geometry of the material, and the
kinematics of the molecular and dusty winds (leading in some cases to the discovery of fast, i.e. 
$\geq100\,{\rm km\,s^{-1}}$ outflows).

In contrast, the polarisation of both the dust continuum and molecular lines
are still poorly studied. The polarimetric information at these longer
wavelengths is primarily used to trace the magnetic field's geometry and to
study anisotropy in the envelope; it also yields information on the dust
grains properties (e.g. nature, size). Submm-to-cm polarimetric investigations are then important tools for the understanding of PPNe.

Dust continuum polarisation is based on the principle of alignment of
non-spherical spinning dust grains with their long axis (and therefore the
polarisation angle) perpendicular to the magnetic
field \citep{Lazarian2003,Lazarian2011}. The linearly polarised emission of
these grains therefore gives a direct image of the magnetic field
distribution/geometry by rotating the polarisation vectors by 90 degrees.
 It is important to notice however that this process is generally valid for simple and constant field geometry at different scales \citep{Matthews2001,Matthews2002}. 

 Former investigations at submillimeter wavelengths based on this mechanism include especially the work of \citet[and references therein]{Hildebrand1984,Hildebrand1996} on star forming regions and molecular clouds.
Only two studies exploiting dust continuum polarisation/alignment, and focusing on evolved stars
(i.e. PNe \& PPNe) have been published and these included only one PPN. In
particular, \citet{Greaves2002} and \citet*{Sabin2007} both targeted CRL 2688
and showed the presence of linearly polarised dust emission at $450\,\micron$
and $850\,\micron$ in the nebular envelope using the submillimeter instrument
SCUBA on the James Clark Maxwell Telescope (JCMT). These studies revealed a
complex magnetic field structure with what appeared to be a combination of
toroidal and poloidal fields.

In the case of molecular lines\footnote{\it excluding masers} the Goldreich-Kylafis
effect \citep{Goldreich1981,Goldreich1982} predicts that the emission from
rotating molecules can be polarised (in the order of a few percent) in the
presence of a magnetic field. The polarisation occurs when anisotropic
radiation alters the molecular magnetic sub-levels. \citet{Morris1985}
proposed another explanation and linked the polarisation to a stellar
radiation field resulting in a preferential rotational direction for the
molecules. The latter hypothesis does not necessarily imply the action of
magnetic fields. However contrary to the case of dust polarisation,
emission-line polarisation vectors can be either parallel or perpendicular to
the magnetic field. The polarisation direction will be strongly affected by
the physical conditions within the objects (e.g. radiation field, hydrogen
number density, magnetic field). Few evolved objects have been investigated
using their polarised molecular lines and the most recent papers 
focused on AGB stars such as IRC+10216 \citep {Girart2012} and IK
Tau \citep{Vlemmings2012}. PPNe have not yet been studied
using this method.

The combination of both dust and molecular polarisation is a valuable tool to determine the magnetic field geometry and anisotropy in PPNe.  We present such a dual polarimetric study on the circumstellar envelopes of two well-known PPNe:  CRL 618 and OH 231.8+4.2.

CRL 618 (also named the Westbrook Nebula) is a $\sim$200 years
old \citep{Kwok1984} PPN located at $\sim$0.9
kpc \citep{Goodrich1991,Sanchez2004}. The nebula is characterised by
two pairs of shocked, rapidly expanding lobes \citep{Balick2013}, belonging to the `multipolar' morphological class
following \citet{Sahai2007}, a central compact HII region and an ancient AGB
halo \citep{Sanchez2002}. CRL 618 is also known for its rich carbonaceous 
dust and
molecular content  which have been extensively
studied at submillimeter, millimeter and centimeter wavelengths from ground based
to space telescopes \citep{Phillips1992, Nakashima2007, Bujarrabal2010, Tafoya2013}.

OH 231.8+4.2 (aka the Rotten Egg Nebula or Calabash Nebula) is a
slightly older PPN \citep[$\simeq$770 yr,][]{Alcolea2001} with a bipolar
morphology as illustrated by its two asymmetrical elongated lobes. Similar to CRL 618, the PPN displays an external round halo. The nebula is located at
$\simeq$1.12 or 1.30 kpc according to \citet{Choi2012} and \citet{Kastner1992}
respectively and shelters a binary star system (Main Sequence A-type + Mira
(QX Pup) following \citealt{Sanchez2004b}b).  OH 231.8+4.2 has a rich molecular
spectrum and an oxygen-rich chemistry, although carbonaceous species such as
HCN, HNC and CS are also seen \citep{Morris1987}.  This PPN has therefore been
widely studied from far-IR to radio wavelengths by
e.g. \citet{Morris1987,Sanchez1997,Alcolea2001}. Its magnetic field has been
studied by \citet{Etoka2009} and \citet{Ferreira2012,LealFerreira2013} using OH
and H$_{2}$O masers, respectively. The latter authors derived a magnetic field
of $\simeq$45\,mG which extrapolated to the stellar surface and assuming a
toroidal field configuration gives a field of $\simeq$1.5--2.0\,G. \citet{Etoka2009} derived a radial field along the outflow. 
We note that maser studies provide very high spatial resolution but sparse sampling, and less information on large scales.

Our submillimeter investigation is organised as follows. In section $\S$2 we describe the observations and data reduction. Section $\S$3 includes the continuum polarisation results for CRL 618 and OH 231.8+4.2 respectively followed by the line polarisation results. Finally the discussion and concluding remarks are presented in sections $\S$4 and $\S$5 respectively.

\section{Observations and data reduction}

We observed CRL 618 and OH 231.8+4.2 with the Submillimeter Array (SMA)\footnote{The Submillimeter Array is a joint project between the Smithsonian Astrophysical Observatory and the Academia Sinica Institute of Astronomy and Astrophysics, and is funded by the Smithsonian Institution and the Academia Sinica.} \citep*{Ho2004} in polarimetric mode \citep{Rao2005}. We used the compact configuration giving a maximum baseline of $\simeq$77\,m. The selected frequency range is divided between the Lower Side Band (LSB) ($\simeq$330--334\,GHz) and the Upper Side Band (USB) ($\simeq$342--346\,GHz). These ranges were chosen to also cover the CO lines in the $J$=3$\rightarrow$2 transition, i.e. $^{13}$CO and $^{12}$CO, at rest frequencies 330.587\,GHz and 345.796\,GHz respectively.

The correlator setup provides a spectral resolution of $\sim$0.8\,MHz (i.e. 0.70\,km\,s$^{-1}$) at 
345.796\,GHz. 3C111 was used as gain calibrator and 3C279 as a bandpass and polarisation calibrator for both objects.
CRL 618 was observed on 2011 November 28 for $\simeq$16 hours (including the calibrators) with a telescope phase center $\alpha_{J2000}={\rm 04^{h} 42^{m} 53\fs64}$, 
$\delta_{J2000}=+36\degr 06\arcmin 53\farcs4$. The weather conditions were excellent with a mean zenith 
$\tau$ of 0.045 at 225\,GHz and stable phases during the observations. OH 231.8+4.2 was observed on 2011 December 27 for $\simeq$14 hours (including the calibrators) with a telescope phase center 
$\alpha_{J2000}={\rm 07^{h} 42^{m} 16\fs83}$,  $\delta_{J2000}=-14\degr 42\arcmin 52\farcs1$. The weather conditions were not optimal due to snow at the summit. The mean $\tau$ was 0.075 (the maximum value reached was 0.13) and the phases were also unstable during the observing run.

The flux, gain and bandpass calibration was performed with the software {\sc mir}\footnote{https://www.cfa.harvard.edu/$\sim$cqi/mircook.html} and then exported to {\sc miriad} \citep{Wright1993,Sault2011} for polarization calibration and imaging. Our data were corrected for polarisation leakage (i.e. instrumental polarisation); we found for both objects consistent leakage terms of $\simeq$5 per cent in the lower sideband and $\simeq$2 per cent in the upper sideband consistent for both objects. We also emphasize that a strong polarization calibrator such as 3C279 coupled with a good coverage of parallactic angle results in an accuracy of 0.1 per cent in instrumental calibration \citep{Marrone2008}.

\section{Results and Analysis}

\begin{figure*}
\begin{center}
\vspace{-1cm}
{\includegraphics[height=15cm,angle=-90]{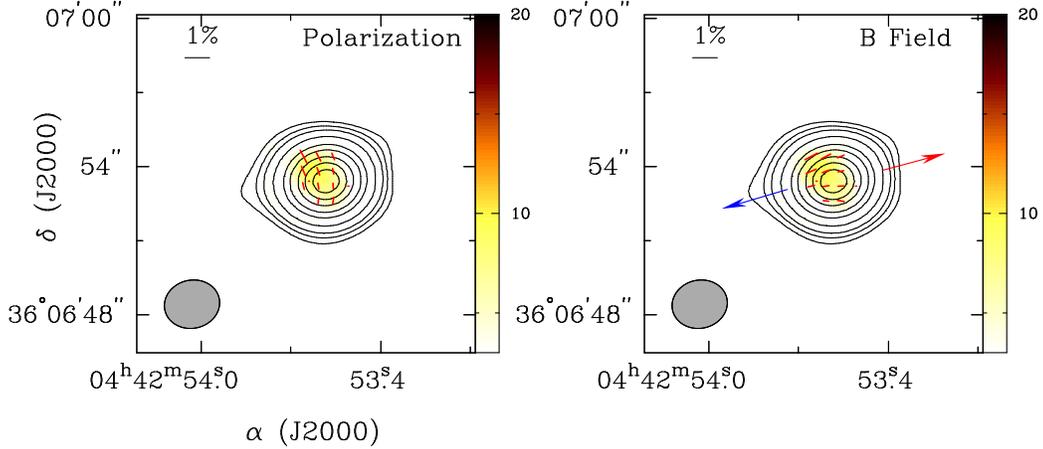}} \vspace{-2cm}
\caption{\label{Maps} Left: Combined (USB+LSB) polarisation map of CRL
618. Right: Combined magnetic field map derived by rotating the polarisation 
vectors  by 90 degrees. The magnetic field vectors show a geometry which can
be linked to a polar field distribution mostly aligned (in the center) with
the outflows indicated by the arrows. The red and blue correspond to the red- and blue-shifted CO outflow (see Fig.\ref{Boutflow} later in this article). In both cases, the black contours which indicate the total dust emission are drawn in steps of 0.02 Jy $\times$ (3,6,10,20,40,60,90,120,150). The color image indicates the polarised intensity with its associated scale bar in Jy/beam on the right. Finally the polarisation vectors are drawn as red segments and the
scale is set to 1 per cent.  The segments are greater than 2.5$\sigma$ detections with the peak value of 4.4$\sigma$. The uncertainty in PA due to thermal noise is about 5 degrees. North is up and East is left.  In all maps the polarisation segments were slightly oversampled when exporting the data from 
{\sc miriad--cgdisp} display tool to the interactive graphics software {\sc wip} \citep{Morgan1995}.  }
\end{center}
\end{figure*}

\begin{figure*}
\begin{center}
\vspace{-1cm}
{\includegraphics[height=15cm,angle=-90]{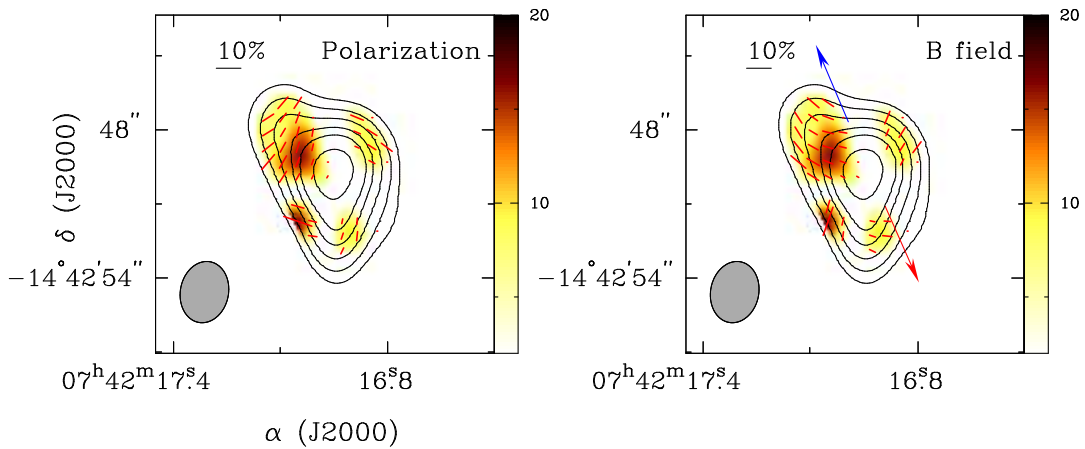}}\vspace{-2cm}
\caption{\label{Maps2} Same as Fig.~\ref{Maps} for OH 231.8+4.2 with Left:
Combined (USB+LSB) polarisation map. The continuum contours are drawn in steps of 
0.02\,Jy\,$\times$\,(3,6,10,15,20,30,40,50,60); Right: Combined magnetic field map with the polarisation vectors rotated by 90 degrees. The field's structure shows an X-shaped (or dipole) configuration. Similarly to CRL 618 the arrows indicate the direction of the (blue and red-shifted CO) outflows. The polarisation vectors are drawn as red segments and the scale is set to 10 per cent. }
\end{center}
\end{figure*}

\begin{figure*}
\begin{center}
\vspace{1cm}
{\includegraphics[height=6cm]{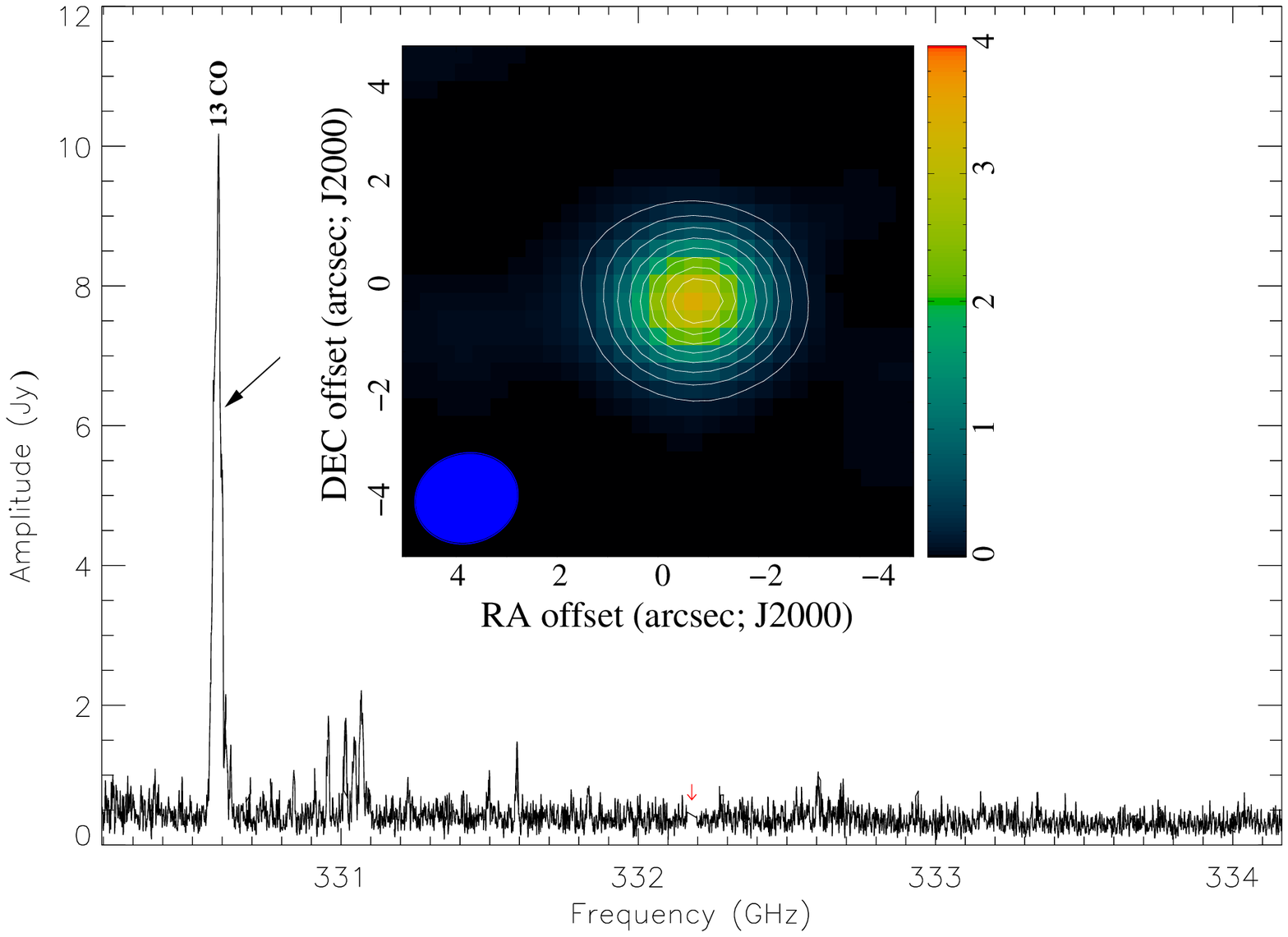}}
{\includegraphics[height=6cm]{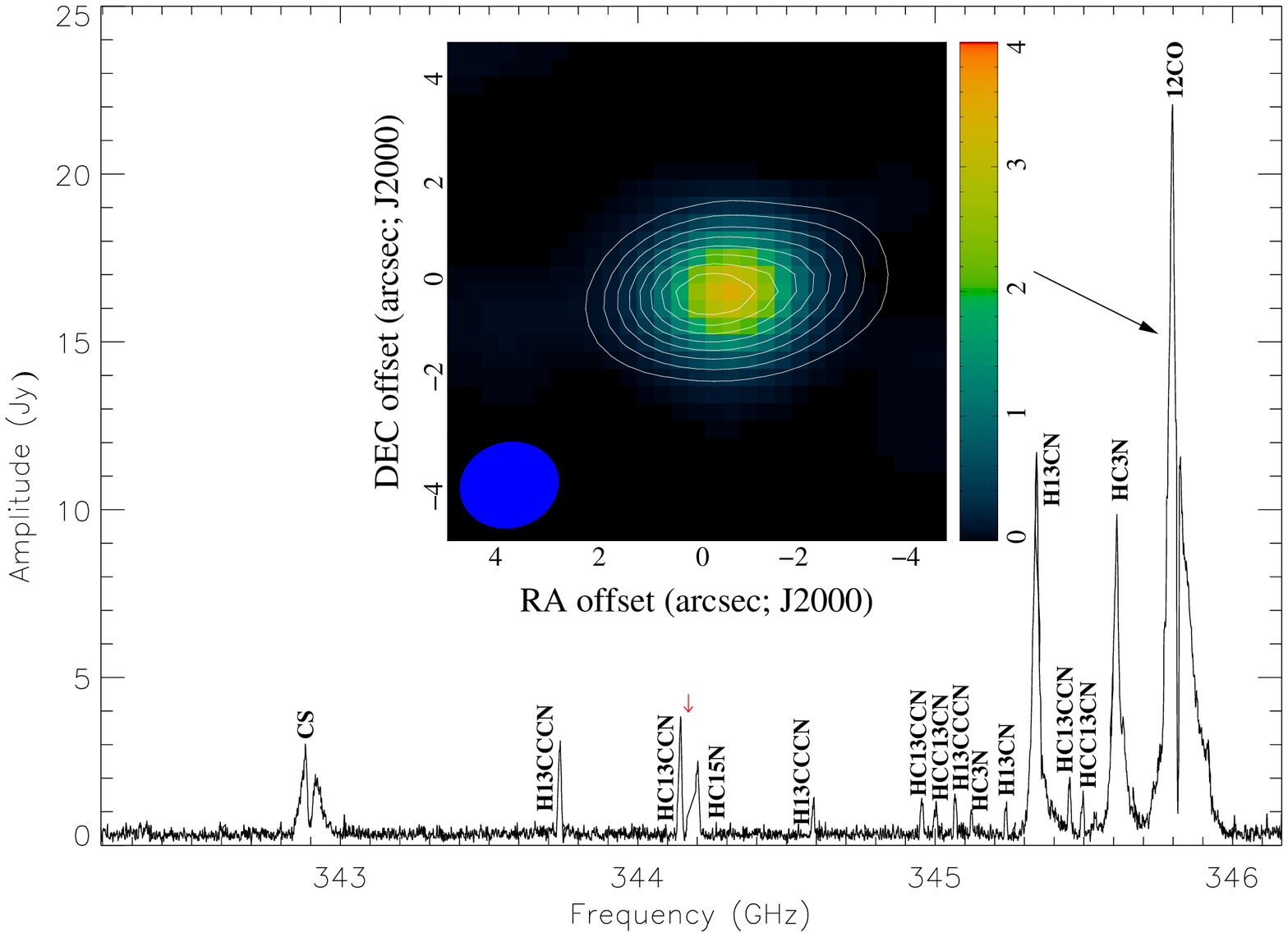}}
\caption{\label{SpectraCRL} LSB (left) and USB (right) visibility amplitude spectra of CRL
618. The emission is averaged over all SMA baselines. The red arrow indicates the 32 MHz coverage gap width. We show for each spectra the geometry of the integrated emission of the CO lines (contours) over the continuum (color) as well as some well identified emission lines. Line splitting is seen in the CS and $^{12}$CO lines. The contours are set as steps of (2,3,4,5,6,7,8,9)\,$\times$\,32.9\,Jy/beam (km s$^{-1}$)$^{-1}$  in the LSB map
and $\times$139.3\,Jy/beam\,(km s$^{-1}$)$^{-1}$ in the USB map.  }
\end{center}
\end{figure*}

\begin{figure*}
\begin{center}
\vspace{1cm}
{\includegraphics[height=6cm]{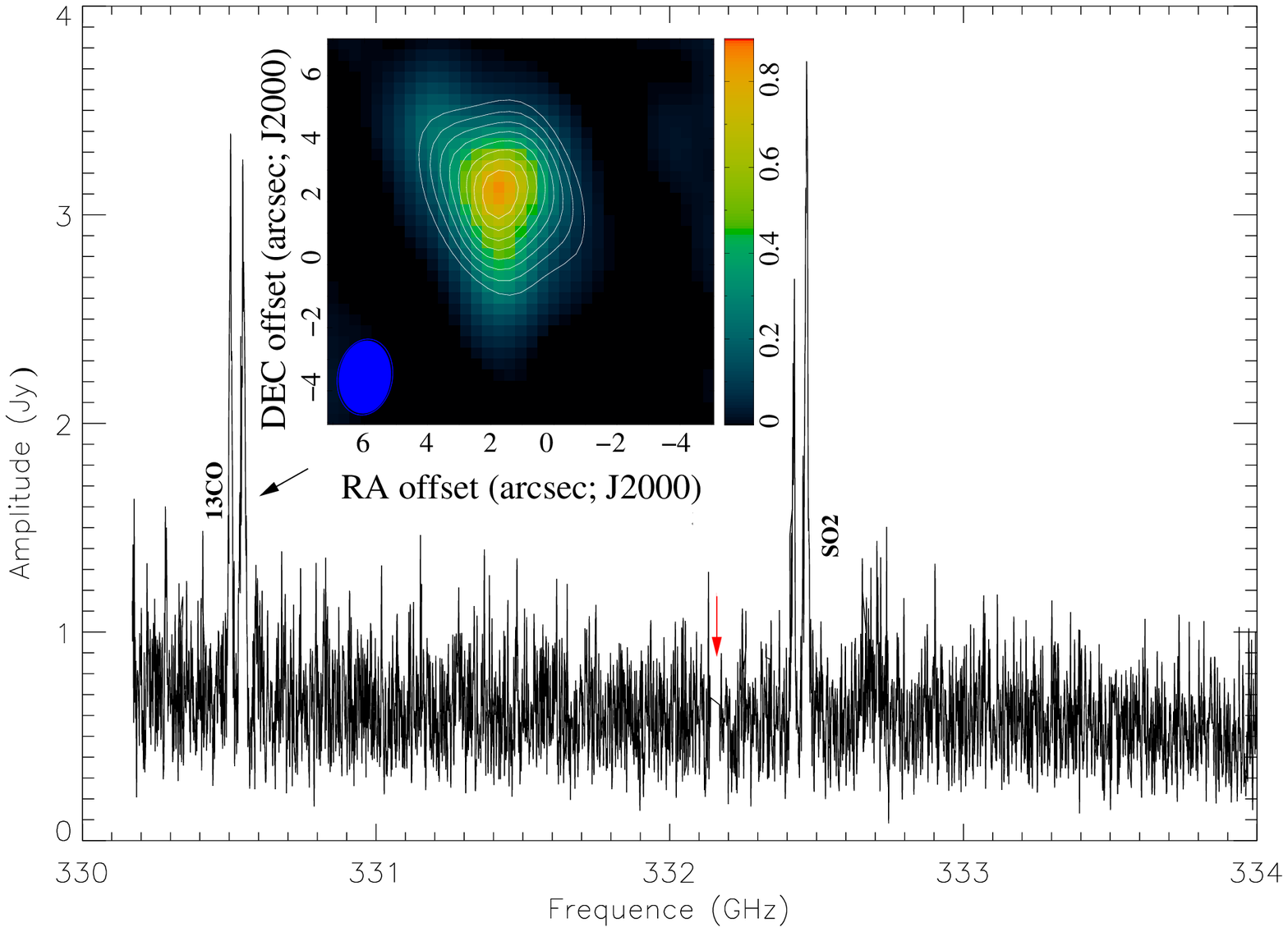}}
{\includegraphics[height=6cm]{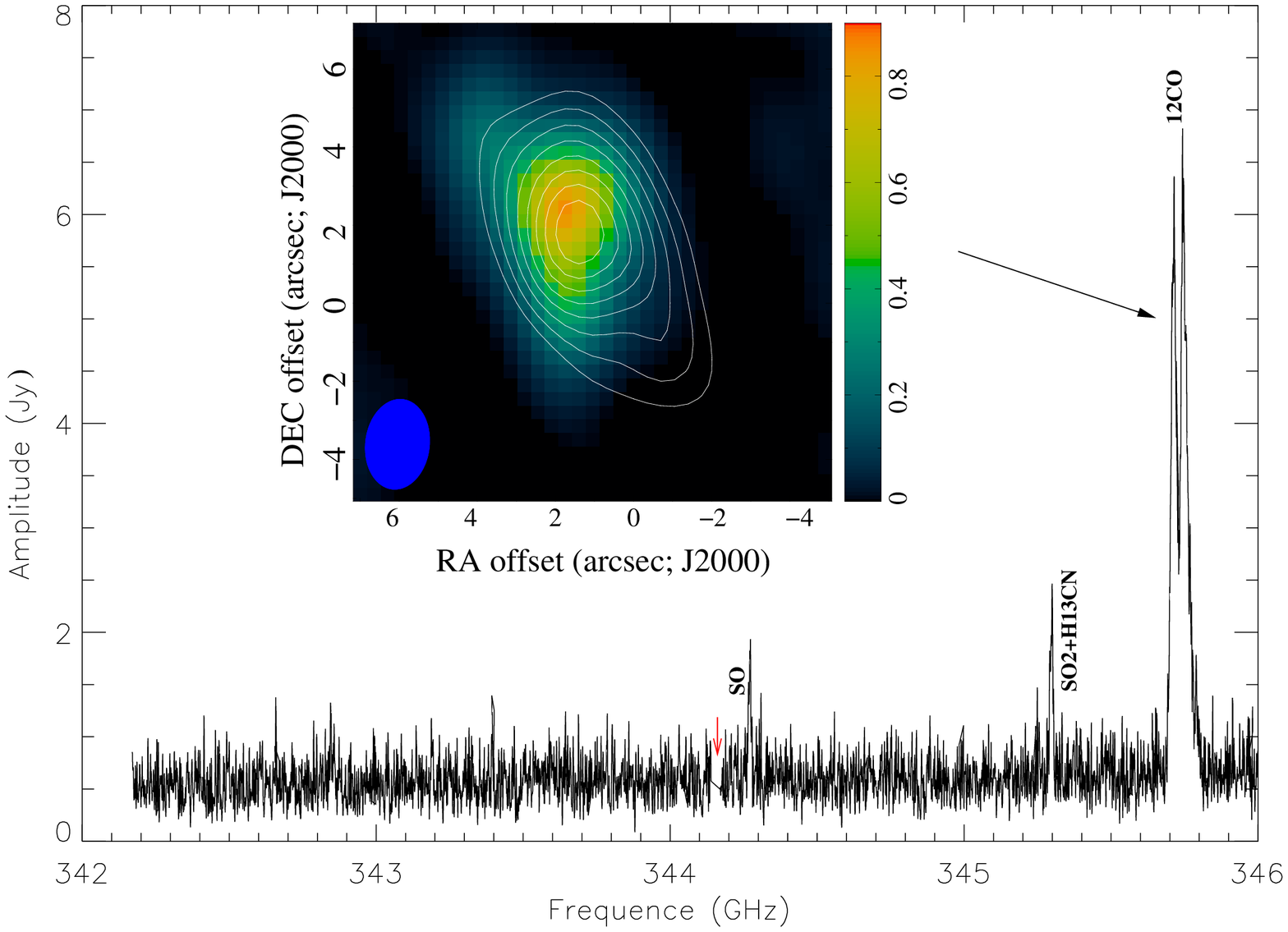}}
\caption{\label{SpectraOH} Same as fig.\ref{SpectraCRL} for OH
231.8+4.2. The spectra are quite noisy (compared to CRL 618 and due to the wet weather) with the LSB being noisier than the USB as expected. The high resolution provided by the SMA allow us to identify different line splittings for the $^{13}$CO, SO$_{2}$ and $^{12}$CO lines (in both objects), which are likely linked to velocity motions as well as the large expansion of the molecular lines. Thus, in the case of the $^{12}$CO(3$\rightarrow$2) line, the splitting is indicating of a fast CO outflow. For each spectra we emphasize the distribution of the integrated emission of the CO lines (contours) over the continuum (color). The contours are set as steps of (2,3,4,5,6,7,8,9)\,$\times$\,27.3\,Jy/beam\,(km s$^{-1}$)$^{-1}$ in the LSB map and $\times$\,39.5\,Jy/beam\,(km s$^{-1}$)$^{-1}$ in the USB map.}
\end{center}
\end{figure*}

The high resolution provided by the SMA
allow us to identify different line splitting in both objects which are likely
linked to velocity motions as well as the large expansion of the $^{12}$CO($J$=3--2) molecular line in CRL 618 indicating a fast CO outflow

\subsection{Continuum polarisation}
\subsubsection{CRL 618}

The continuum was carefully selected and subtracted from both the LSB and USB avoiding the inclusion of emission lines such as the relatively strong
CO($J$=3--2), CS($J$=7--6), H$^{13}$CN($J$=4--3), HC$_3$N($J$=38--37) by visually examining the visibility amplitude spectra. 
The final maps of the thermal continuum emission in CRL 618 (relative to dust polarisation and magnetic fields) were made combining the line-free channels of the USB and LSB. To obtain the
best sensitivity for the Stokes $I$ (total intensity) and Stokes $Q$ and $U$ (linear
polarisation), we used a robust weighting (in order to minimise the noise
level) resulting in a synthesised beam of 2.2\,$\times$\,1.9 arcsec with a
position angle of $-77\fdg6$. The measured rms noise levels,
$\sigma_{I}$=19.8 mJy/beam and $\sigma_{Q,U}$=2.2 mJy/beam, defined the zero
levels we applied to establish the polarisation maps. The polarised intensity
and percentage of polarisation ($I_{\rm P}$ and $P(\%)$ respectively) are defined as
the following: $I^{2}_{\rm P}=Q^{2}+ U^{2} - \sigma^{2}_{Q,U}$ and $P(\%)=I_{\rm P} /
I$ (with $I$ the total intensity). 

The continuum dust emission was detected over an area of approximatively
5.4\,$\times$4.6\,arcsec centred at the coordinates $\alpha={\rm 04^{h} 42^{m}
53\fs582}$, $\delta=+36\degr06\arcmin53\farcs40$ (Fig.~\ref{HST}-Top). The
continuum emission is slightly elongated east-west and this is consistent with
the results by \citet{Pintado1993} (although at higher angular
resolution). The measured peak intensity, located in the dark lane of the PPN,
is 3.4\,Jy/beam with a mean of 1.2\,Jy/beam over the full area of the continuum
emission.

Linear polarisation is detected above $\simeq$3$\sigma$ across the source. The
polarimetric information obtained reveals a peak of 9.6\,mJy/beam 
(at $\alpha={\rm 04^{h} 42^{m} 53\fs621}$, $\delta=+36\degr06\arcmin53\farcs79$) 
and a mean
polarised emission of 7.0\,mJy/beam which correspond to 4.4$\sigma$ and
3.2$\sigma$ detections respectively. The peak continuum intensity is not
coincident with the peak fractional polarisation $P(\%)=0.7$. However, the
mean polarisation over the whole structure is quite low with $P(\%)=0.3$. 
Those values depend on the assumed zero level (derived from the noise
level), but we do not expect it to reach values greater than $\sim$1 per cent.
The polarisation vectors can be divided in two main sets which overall form a
slightly curved pattern opening towards the East (Fig.\ref{Maps}-Left). The
first set is linked to the upper (northern) half of the polarised continuum
and shows a mean polarisation angle of $\simeq$22\degr; the second set is
linked to the lower (southern) half of the polarised continuum and shows a
mean polarisation angle of $\simeq$--10\degr. We however have to be careful while discussing the variation in polarization angle as the polarization emission is not spatially resolved, as shown by the size the of synthesised beam. Assuming that the magnetic field direction is given by rotating the dust polarisation vectors, in Fig.\ref{Maps}-Left, by 90 degrees, we obtain the magnetic map in
Fig.\ref{Maps}-Right which we discuss later in this article ($\S$4).

\begin{figure*}
\begin{center}
\vspace{1cm}
{\includegraphics[height=13.5cm,angle=-90]{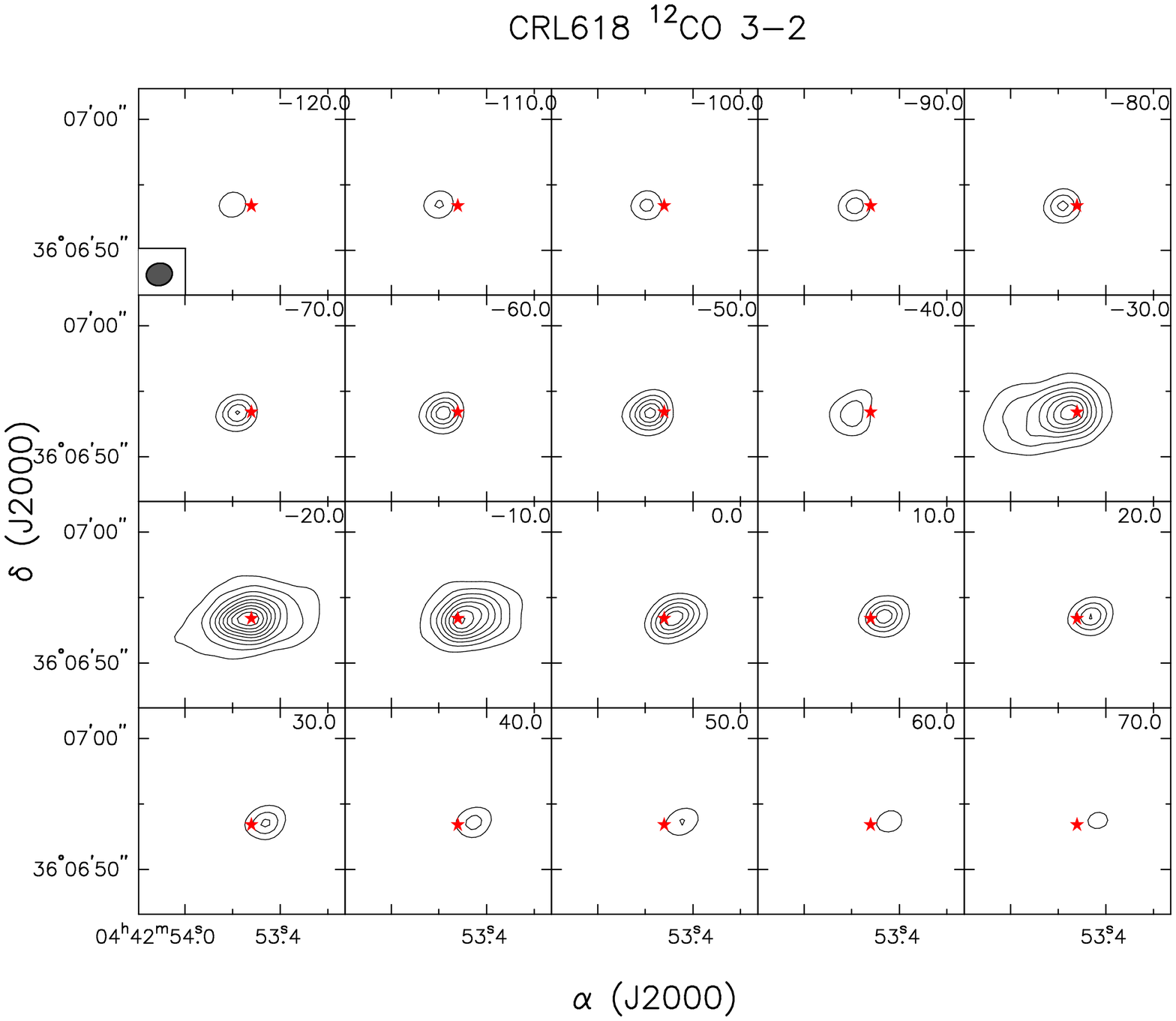}}
{\includegraphics[height=13cm,angle=-90]{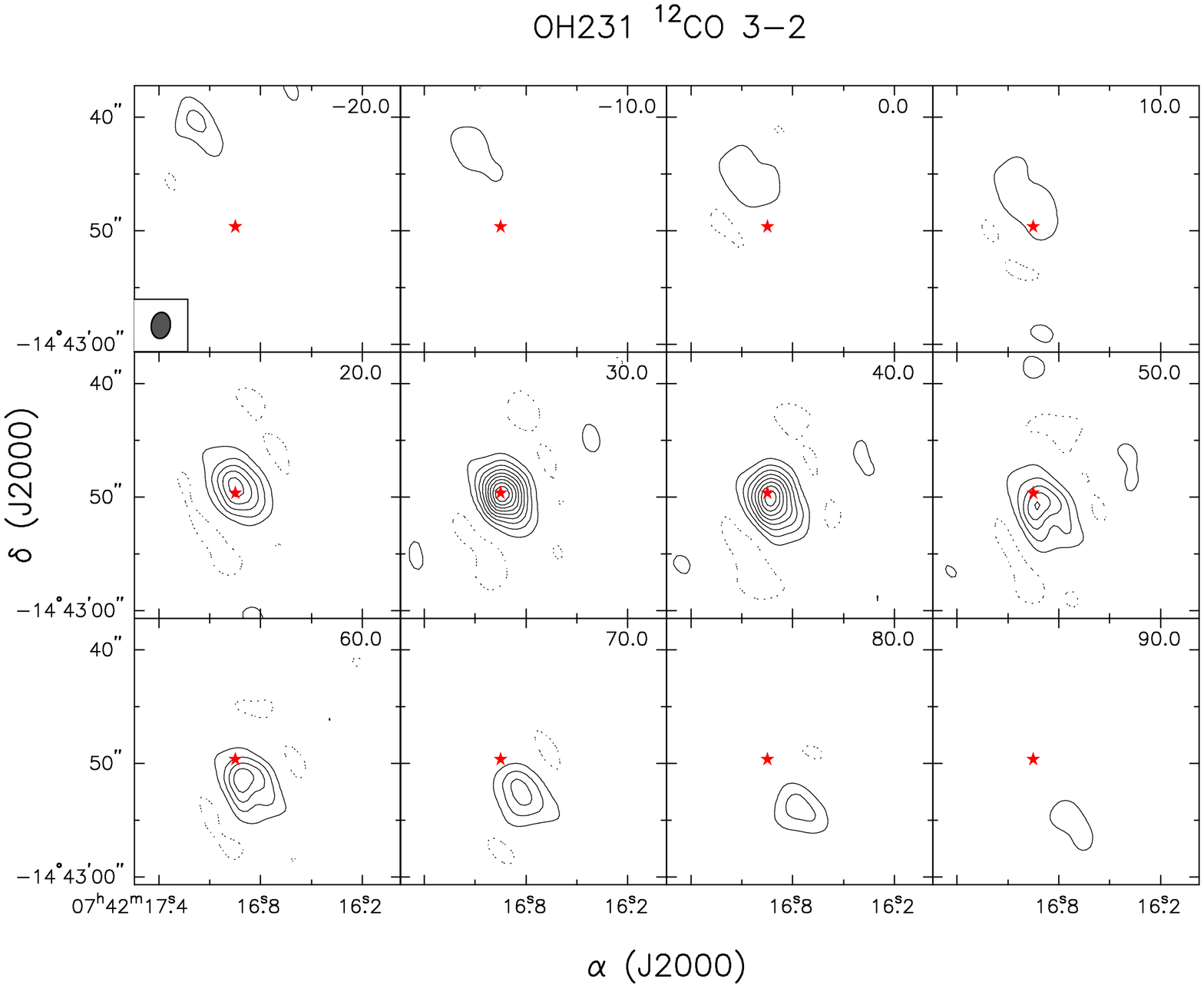}}
\caption{\label{channels} $^{12}$CO channel maps for CRL 618 (Top) and OH 231.8+4.2 (bottom). The contours are set at 10 per cent of the peak. In the channel maps, the peak of the continuum is represented by the star symbol and the velocity in km\,s$^{-1}$ is indicated in the top-right corner. Both sets of maps indicate the presence of high velocity gas in the PPNe.}
\end{center}
\end{figure*}

\begin{figure*}
\begin{center}
{\includegraphics[height=9.5cm,angle=-90]{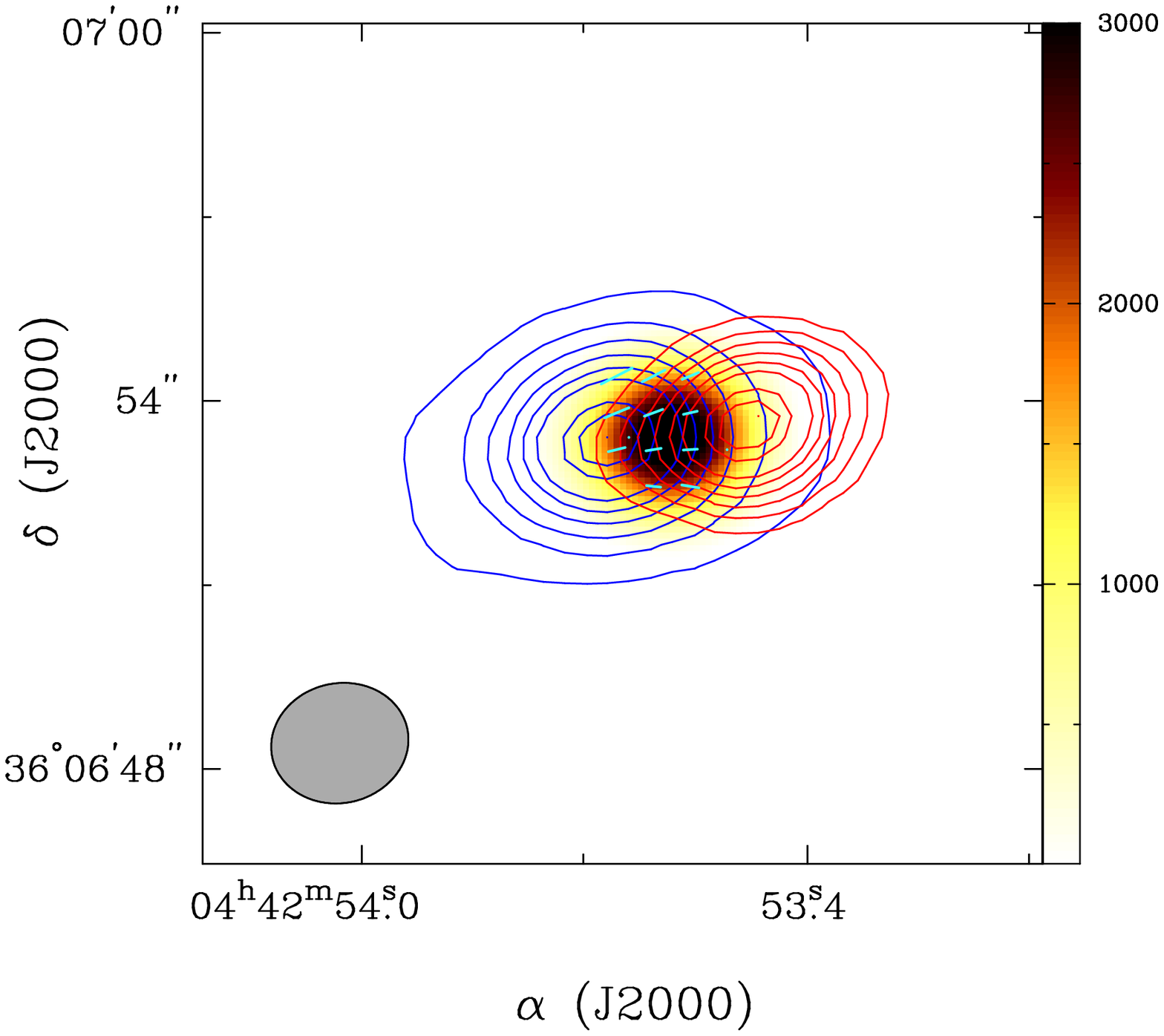}} \hspace{-2cm}
{\includegraphics[height=9.5cm,angle=-90]{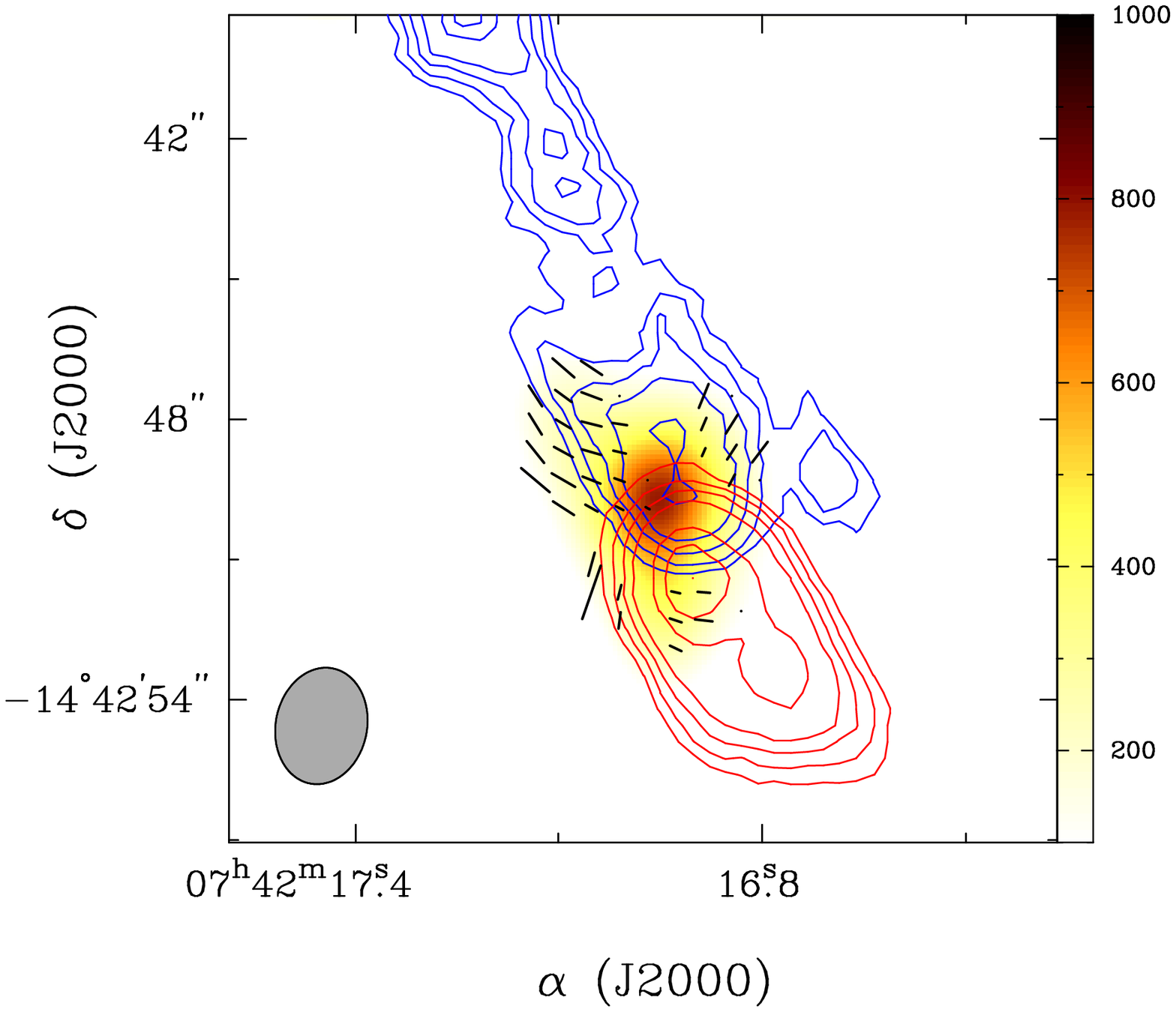}}
\caption{\label{Boutflow} Distribution of the $^{12}$CO($J$=3$\rightarrow$2) outflow {\it vs} the magnetic field. {\it Left panel}: The blue and red-shifted lobes of the $^{12}$CO emission in CRL 618 are drawn in contours starting at 10 per cent and in steps of 10 per cent of the peak 
(blue peak\,=\,828\,Jy\,$\times$\,km\,s$^{-1}$; red peak\,=\,399\,Jy\,$\times$\,km\,s$^{-1}$). The CO outflows are integrated over a velocity range of --110 to --20\,km\,s$^{-1}$ for the blue-shifted lobe and +5 to 
+90\,km\,s$^{-1}$ for the red-shifted lobe. We can therefore see the integrated emission intensity for each element. The two lobes are aligned with the continuum (Stokes $I$) emission (color scale in mJy) and a good agreement is also observed with the magnetic vectors distribution (cyan segments). {\it Right panel}: Same as the left panel but for OH 231.8+4.2. In this case, the $^{12}$CO emission contours start at 30 per cent and in steps of 10 per cent of the peak 
(blue peak\,=\,100\,Jy\,$\times$\,km\,s$^{-1}$; red peak\,=\,158\,Jy\,$\times$\,km\,s$^{-1}$).
The CO outflows are integrated over a velocity range of --60 to +15\,km\,s$^{-1}$ for the blue-shifted 
lobe and +50 to +110\,km\,s$^{-1}$ for the red-shifted lobe. We observed an elongated blue shifted CO emission and a good correlation between the magnetic vectors (black segments) and the general outflow distributions.}
\end{center}
\end{figure*}

\subsubsection{OH 231.8+4.2}

The robust weighting applied resulted in a synthesized beam of 
$\simeq$2.5$\times$1.9\,arcsec with a position angle of --11\fdg1. The rms
noise derived for the Stokes $I$, $Q$, and $U$ was $\sigma_{I}$=20.5\,mJy/beam, and
$\sigma_{Q,U}$=4\,mJy/beam. The thermal emission from the continuum appears as
an elongated distribution extending over an area of
$\simeq$7.3$\times$5.7\,arcsec$^2$ and centred at the coordinates 
$\alpha={\rm 07^{h} 42^{m} 16\fs979}$, $\delta=-14\degr42\arcmin49\farcs70$
(Fig.~\ref{HST}-Bottom). The peak intensity (located at the coordinates
$\alpha={\rm 07^{h} 42^{m} 16\fs955}$, $\delta=-14\degr42\arcmin49\farcs77$)
is 0.78\,Jy/beam with a mean of 0.31\,Jy/beam over the whole area.  The
polarised structure in OH 231.8+4.2 strongly differs from that of CRL 618 as it
appears fragmented into four parts (Fig.~\ref{Maps2}-Left). The largest
component ($\simeq$2$\times$1.5\,arcsec) is located in the north-east
section and shows a peak polarised intensity of 16\,mJy/beam 
($\simeq$4$\sigma$; which is not exactly coincident with the peak of Stokes $I$) 
and a mean polarised intensity of 11\,mJy/beam ($\simeq$2.7$\sigma$) is 
measured over this polarised area. The south-east component shows a similar 
polarised intensity of $\simeq$16\,mJy/beam (mean of 11\,mJy/beam) and this 
is the location of the strong peak percentage of polarisation with a value of
$\simeq$15.6 per cent (the mean percentage is 6.7 and 4.3 per cent over the 
whole structure). The north-west and south-west components both show polarised
intensities of $\sim$10\,mJy/beam with a mean of $\sim$9\,mJy/beam. We 
highlight a slight east-west gradient in the polarised intensity
distribution. The intervening positions between the four regions do not have polarization detection above 2.5$\sigma$ (which is the cutoff in Stokes $Q$ and $U$ when computing polarization angles for OH231). The repartition of the electric vectors is also quite
interesting as each component presents not only well organised patterns but
also a main position angle (PA) with values and signs different from the
immediate neighbor. Indeed, the north-east and north-west spots have global
PAs of approximately --34$\pm$18\degr and +57$\pm$3\degr,
respectively, the south-east and south-west spots show global PAs of
+70$\pm$3\degr and --8$\pm$5\degr, respectively. The derived
magnetic field map (Fig.~\ref{Maps2}-Right) is different than that of CRL 618 
due to the fragmented detected polarised emission ({\it see discussion in section
$\S$4 }).

\subsection{Spectral line polarisation}

In an attempt to detect molecular line polarisation we also targeted the
strongest molecular lines present in both objects: $^{13}$CO($J$=3$\rightarrow$2) in the
LSB and $^{12}$CO($J$=3$\rightarrow$2) in the USB. The SMA spectra are shown in
Fig.~\ref{SpectraCRL} and Fig.~\ref{SpectraOH} for CRL 618 and OH 231.8+4.2
respectively. We note that recently, \citet{Lee2013a} presented a higher resolution ($\sim$0.3 
and $\sim$0.5 arcsec)
spectrum as well as high resolution maps, of CRL 618 at 345\,GHz also using the
SMA, although they do not cover our full spectral range.

The noise in the (Stokes $Q$ and $U$) emission line is higher than that of the
continuum, requiring a higher zero-level to assert the presence of
polarisation. The analysis of those lines did not return any detection above
3$\sigma$ and therefore any conclusive signs of polarisation. However the percentage 
of polarisation is not expected to be high and we estimate an upper limit between 5 and 
10 per cent. Indeed, as previously mentioned, similar investigations by \citet{Girart2012} and \citet{Vlemmings2012} obtained detectable polarisation at sigma levels
greater than $\sim$4--5 in the AGB stars IRC+10216 and IK Tau. In those cases
the maximum polarisation fraction varied from $\simeq$2 up to $\simeq$13 per cent
for the different lines observed in both objects (SiS\,19--18, CS\,7--6, 
CO\,3--2, CO\,2--1, and SiO\,5--4). 
%(for $^{12}$CO the rms is $\sim$46 mJy/Beam for CRL 618 and 142 mJy/Beam for OH 231.8+4.2)

\section{Discussion}

The SMA data obtained indicates the presence of a polarised continuum at
submillimeter wavelengths in the
proto-planetary nebulae CRL 618 and OH 231.8+4.2. Several observations can be
made regarding the dust polarisation in these objects and the magnetic field
geometry. 

\subsection{Grain dust properties}

We observed that OH 231.8+4.2 is more polarised on average, in terms of degree
of polarisation, than CRL 618 despite its non-uniform coverage. Thus, the mean
degree of polarisation in CRL 618 is 0.3 and 4.3 per cent for OH 231.8+4.2. Both
objects differ in their chemistry, and \citet*{Sabin2007} have shown that in their sample, and at
the same wavelength, the mean degree of polarisation of the oxygen-rich
PNe/PPNe is greater than that of the carbon-rich PNe/PPNe. The two new
nebulae studied here also follow this trend. This supports the idea that the
grain alignment efficiency is higher in O-rich than in C-rich envelopes and
therefore dust chemistry plays a role not only in the polarisation process but
also in our interpretation of magnetic field distribution.

Carbon grains have generally a smaller size than silicate grains and may not
produce significant polarisation\footnote{The alignment of polycyclic aromatic
hydrocarbons (PAH) has been discussed by \citet{Lazarian2000}
and \citet{Lazarian2003} and they concluded that the polarisation and
alignment of those small species would be marginal or partial}. Also, it would
be easier or more likely for larger grains to host superparamagnetic
particles \citep{Jones1967,Mathis1986,Kim1993}, such as iron or magnetite,
which would cause a higher level of polarisation and a more efficient dust
grain alignment by the magnetic field.  The polarised intensity appears to be
sensitive to the grain shape as well \citep{Kim1995} and most of the dust
alignment theories require either elongated (e.g. oblate/prolate spheroids) or
irregular grains \citep{Lazarian2003}.  Therefore one would expect that the
presence in the disk of OH 231.8+4.2 of larger grains with iron inclusions,
crystalline olivine  (MgFeSiO$_{4}$), crystalline enstatite
(MgSiO$_{3}$), and H$_{2}$O crystalline ices \citep[via
modelling]{Maldoni2004}, would produce a greater polarisation than CRL 618,
which presents carbonaceous dust (mostly amorphous carbon,
see \citealt{Lequeux1990}).

Unfortunately a detailed study of the dust grain
properties in those objects  is still
missing.

\subsection{Depolarisation effect}

An anti-correlation is seen between the fractional polarisation and the total
intensity in both objects. Indeed, $P(\%)$ tends to decrease from the
edges to the center of the emission area, opposite to Stokes $I$. This
`depolarisation effect' is not new and has often been observed in dark clouds,
molecular clouds and star forming regions \citep[e.g.][]{Chen2012} but also,
and most relevant here, in PNe and
Post-AGBs \citep*{Sabin2007,Greaves2002}. This pattern can be associated with
multiple phenomena such as the effect of beam smearing, a complex magnetic
field distribution \citep{Matthews2001,Greaves2002}, and a loss of efficiency in
the grain alignment \citep{Lazarian1997} due to molecular collisions or the
variation of the density which would alter the structure of the dust grains
and therefore the degree of alignment \citep{Cho2005}.

\subsection{Dust distribution}

As previously mentioned the polarisation indicates the location of aligned
dust. In both PPNe this polarised dust is circumscribed inside the optical
dark lane.

In CRL 618 the polarised grains are roughly perpendicular to the direction of
the ionised outflows (Fig.~\ref{Maps}-Left). The emission, expanding over
$\simeq$3.8\,arcsec$^{2}$, might be related to the dusty torus/ring or to the
central bipolar compact \mbox{H\,{\sc ii}} region (\citealt[]{Kwok1984}\footnote{Note that they
found a rather small radio core structure of $\sim$0.3\,arcsec major
axis}, \citealt[see their Fig.6]{Sanchez2004}). 
%The small polarised spot located in the west indicates a PA slightly coincident with the equatorial plane of CRL 618. This polarised dust distribution is reminiscent of CRL 2688 where directions of alignment along the outflows and the equator were found at 850$\micron$ and 450$\micron$ \citep*{Sabin2007}. But in  the current data this is a marginal structure which would require confirmation.\\

In OH 231.8+4.2 the polarisation vectors, which are divided into four groups,
indicate different polarisation angles and seem to draw the contours of a
dusty region (if we link them all) which might also delimit a ring or torus of
$\simeq$3.6\,arcsec radius well centred on the equator of the PPN
(Fig.~\ref{Maps2}-Left). This distribution appears consistent with the ring-like
structure described by the polarisation vectors associated to the OH masers in \citep{Etoka2009}.\\

\subsection{Magnetic field distribution}

\begin{figure*}
\begin{center}
\vspace{1cm}
{\includegraphics[height=5cm]{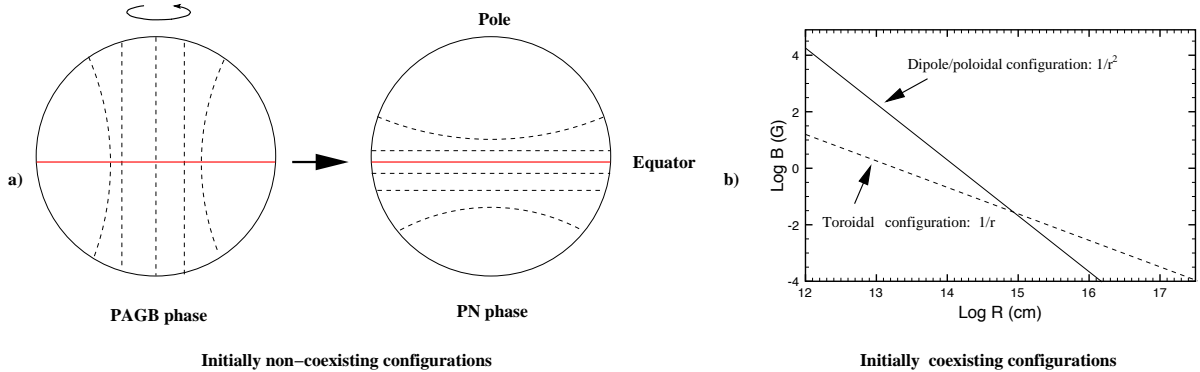}}
\caption{\label{Bdistri}  Sketches illustrating the possible evolutionary scenarios of the magnetic field in evolved stars.  On the left, we show the full change in magnetic configuration i.e. from a dipole/poloidal to a toroidal (via rotation). The dashed lines indicate the magnetic field lines. Detailed toroidal, dipole and quadrupole magnetic field geometries (from modelling) have also been presented by \citet{Padovani2012} and by \citet{Matt2000}. On the right, we present an adaptation of the graphic by \citet{Vlemmings2011} showing the variation of the magnetic field strength in function of the stellar radius and according to the magnetic field configuration (i.e. dipole/polar and toroidal). Assuming that both configurations coexist from the beginning, it is clear that the dipole field will disappear first and faster, leaving a dominant toroidal field. }
\end{center}
\end{figure*}
                  
The magnetic field in CRL 618 is globally aligned with the pairs of ionised
outflows (indicated by the arrows in Fig.~\ref{Maps}-Right). Indeed the
polarisation vectors of the main structure show a mean PA of
$\simeq$96\degr while the PPN has a mean PA along the lobes of
$\simeq$94\degr. The orientation of the magnetic field is mostly at 90 degree
to the equatorial plane of CRL 618. This configuration is pretty
similar to the structure of CRL 2688, which is another carbon rich PPN with
exactly the same optical quadrupolar morphology. CRL 618 therefore shows a well
defined polar magnetic field. The magnetic field appears to be  well
organised. 

The field's distribution in OH 231.8+4.2 is particularly patchy. The eastern
side of the PPN (also showing the higher polarisation intensity) presents in
each small area a very well organised {\it curved} field directed away from
the equatorial plane (Fig.~\ref{Maps2}-Right). The magnetic vectors describe at first sight an X-shape
which would then associate the magnetic field in OH 231.8+4.2 with a dipole
configuration (see model by \citealt{Padovani2012}). 
However, we cannot discard another interpretation that is the low inclination respected to the equatorial plane of some vectors belonging to the north-west and south-east blobs (and to some extent the south-west blob), might indicate the presence of a partial (and not strongly delineated) toroidal field. OH 231.8+4.2 would in this case show a dual configuration such as the PPN CRL 2688.

%The deduced magnetic field distribution can lead to different interpretations regarding the general evolution of the field in late type low-intermediate mass stars in general. We present here two possible scenarios.

It is interesting to notice that neither of these two high mass nebulae show a well defined and constrained toroidal equatorial magnetic field such those seen in the planetary nebulae NGC 6537, NGC 6302 and NGC 7027, which in turn show no poloidal field \citep*{Sabin2007}. But they coincide more with the magnetic distribution of CRL 2688 where the field is mainly aligned with the polar direction. CRL 618, OH 231.8+4.2 and CRL 2688 being younger than the `NGC group', we may see here a correlation with the evolutionary stage of the nebulae. This would then indicate that there is either a tendency for the distribution of the magnetic field to evolved from a single dipole/poloidal configuration to a toroidal one (passing through a phase of dual configuration) while the nebulae evolve (Fig.~\ref{Bdistri}-Case a). This type of transition generally occurs via the mechanism of rotation. An alternative is that both configurations might exist simultaneously (Fig.~\ref{Bdistri}-Case b). In this case we postulate that the initially weak toroidal field  (as it is often the case in slow rotator objects) could coexist with a stronger poloidal one. As the nebula expands, the later declines quickly, as $r^{-2}$, respective to the toroidal field which begins to dominate because it declines as $r^{-1}$. This would explain why no poloidal configuration is seen in more evolved nebulae e.g. PNe (see also \citealt{Gardiner2001}).

A greater statistical sample and a multi-scale analysis of the magnetic field distribution are still needed to answer the `evolutionary problem'.

So how do magnetic fields account for the bipolar or multipolar shapes observed?
 As we observed non-spherical geometry before the occurrence of the toroidal fields the latter cannot be the main dynamically shaping agent, i.e. with enough energy to constrain the material in the equatorial plane and launch collimated outflows. 
However it has been theoretically claimed that dipole magnetic fields can as well lead to non-spherical geometry. For instance, \citet{Matt2000} have shown that a dipole field could create a dense equatorial torus and collimate the winds (with no need of a companion star). The newly created aspherical geometry would then be conserved and/or enhanced with the evolution towards a toroidal field.

Taking into account all the recent works arguing for the major role played by close binary systems in the shaping of PPNe/PNe  \citep[e.g.][]{Soker2004,Nordhaus2008,Douchin2013,Tocknell2013}, a plausible scheme would involve a rotational effect induced by binary interaction which then would have some dynamical effects on the formation and evolution of the magnetic field and on the objects' morphologies.

\begin{table*}
\begin{center}
%\small\addtolength{\tabcolsep}{-4pt}
\caption[]{\label{Maser} Summary of masers study in CRL 618 and OH 231.8+4.2}
\begin{tabular}{|l|l|l|l|l|l|}
\hline
PPN     &     Maser Type  &   B-Field strength  &  Field configuration &  B-Field extrapolation  & Reference  \\ 
\hline
\hline
CRL 618 &   CN ($N$=1--0)   &   0.9 G(*) &  Toroidal($\dagger$): $r^{-1}$  &  200 G(*) at 8 R$_\star$&   \citet{Herpin2009}\\ 
        &                 &                      &            &  2 kG(*) at 1 R$_\star$&  \\
\hline
\hline
OH 231.8+4.2 &  OH              &    -    & Poloidal \& Toroidal     & - & \citet{Etoka2009} \\
           &    H$_{2}$O  & 45 mG   &   Toroidal($\dagger$): $r^{-1}$   &    $\simeq$ 2.5 G at 1 R$_\star$ &\citet{Ferreira2012}  \\
\hline
\end{tabular}
\begin{minipage}{15cm}
\begin{flushleft}
(*) Upper limit.($\dagger$) Assumed field configuration. Note: Contrary to the other studies where Zeeman splitting was used to measure the field's strength, \citet{Etoka2009} analysed the distribution of the 1667 MHz OH masers and observed, similar to our study, a well organised magnetic field with a partial orientation of the field along the outflows.
\end{flushleft}
\end{minipage}
\end{center}
\end{table*}

\subsubsection{Molecular outflows and launching mechanism}

The study of the emission lines in both PPNe can help us to investigate the relationship between the outflows and the magnetic field. In both objects we note the relatively strong and high velocity of the CO 
outflows\footnote{See also the recent article by \citet{Lee2013b} for CRL 618.} and as the 
$^{12}$CO($J$=3$\rightarrow$2) line is the most significant in terms of intensity we will then consider that the molecular outflows are mainly supported by this emission line. Fig.~\ref{SpectraCRL} and 
Fig.~\ref{channels}-Top show the CO molecular outflows for CRL 618 while Fig.~\ref{SpectraOH} and 
Fig.~\ref{channels}-Bottom show the same but for OH 231.8+4.2.

A good positional correlation between the continuum area and the $^{12}$CO emission in CRL 618 is seen in Fig.~\ref{SpectraCRL}. The comparison of the direction of the well organised magnetic field vectors with that of the overall CO molecular outflows in Fig.~\ref{Boutflow}-Left, shows that they are well aligned.

Concerning OH 231.8+4.2, Fig.~\ref{SpectraOH} indicates a slight misalignment of the continuum area with respect to the $^{12}$CO emission distribution of $\sim$30\degr. Fig.~\ref{Boutflow}-Right shows that the 
X-shaped (or pinched waist) structure of the magnetic field in OH 231.8+4.2, mostly the northern and eastern segments, encompasses quite well the blue and red-shifted lobes of the $^{12}$CO emission. The magnetic vectors in OH 231.8+4.2 are globally parallel to the CO molecular distribution and trace quite well the outer contours of the red- and blue-shifted CO lobes.

In both cases the alignment between the major axis of the molecular outflows and the ordered magnetic field suggests not only a dynamically important field at small scale but could also indicate the presence of a magnetic launching mechanism of these outflows. Our observations are concordant with the theoretical predictions by \citet{Blackman2001} relative to the presence of a poloidal geometry of the magnetic field at small distance from the central star and its role in the launching of the flows in proto Planetary Nebulae. Recently, \citet{Perez2013} have also reported the radio observation of a magnetically collimated outflow/jet shaping the post-AGB star IRAS 15445-5449. However we cannot discard the possibility of a field dragged by the powerful outflows, particularly in the case of OH 231.8+4.2. Investigations at larger depths, i.e. closer to the central star, are however needed to fully assert the hypothesis outflow launching theory.

\subsection{Magnetic field strength}

\citet[hereafter CF]{Chandra1953} described a method to derive the magnetic field strength based on the dispersion of the polarisation angles, the gas density and rms velocity.

\begin{equation}
B_{\rm POS}= \sqrt{4\pi\rho} ~\frac{\sigma_V}{\sigma_\phi}
\end{equation}

\noindent with $B_{\rm POS}$ the field in the plane of the sky, $\rho$ the gas density, $\sigma_V$ the velocity dispersion, and $\sigma_\phi$ the dispersion in polarisation position angles.
This method has been widely used to estimate the field strength in molecular clouds and star forming regions and has been adjusted over time to give a more accurate representation of this field \citep{Ostriker2001,Falceta2008,Koch2012}. However as it stands the CF method can hardly be applied to our evolved nebulae. The method is based on the assumption that the magnetic field dispersion is Alfvenic but its application in a non-quiescent environment i.e. where turbulences/perturbations occur is problematic. Indeed, although our nebulae could show signs of locally dominant Alfv\'en waves' contribution \citep*{Sabin2007}, they are subject to ionisation and collision processes among other turbulent phenomena, which are likely to alter the vector polarisations and then the PA dispersion. In conclusion the CF method in its actual form is not adapted to account for the magnetic field strength in our type of nebular media.
 
An estimate of the field strength can however be obtained via maser observations and Zeeman splitting analysis. Table~\ref{Maser} summarises the different studies using this method related to CRL 618 and OH 231.8+4.2.  It is worth mentioning that these masers are located in the outer layers of the envelopes and are therefore not very reliable for a global field estimation.

%\citet{Herpin2009} studied the Zeeman splitting of the CN
%\(N=1--0) line (between 113.144 GHz and 113.509 GHz) and could only derive an
%\upper limit for CRL 618 of 0.9 G when scaled  to 8 R$\star$ (where the SiO masers in O-rich nebulae are located) gives an upper limit
%\of 200 G and 2 kG at 1 R$\star$. In all cases this follows a toroidal $1/r$ law
%\for the field strength evolution. But similar to OH masers, CN masers are
%\located in the outer layers of the envelope ($\log R_\star$= $\sim$16--17 cm)
%\and are therefore not very reliable for a global field estimation. \\ 

%\In the
%\case of OH 231.8+4.2, as previously mentioned, \citet{Etoka2009} analysed the
%\1667 MHz OH masers and observed, similar to our study, a well organised
%\magnetic field with a partial orientation of the field along the outflows. But
%\the authors also mentioned the presence of a torus at the equator. More
%\recently \citet{Ferreira2012} conducted an analysis of 30 H$_{2}$O masers
%\detected in this PPN and which also appear to be dragged by the outflows. Only
%\three masers were linearly polarised but in those cases the associated
%\magnetic field direction could not be derived (as based on the maser theory,
%\the polarisation vectors can be either parallel or perpendicular to the field)
%\nor could a global map of the field in this area be established. However, the
%\Zeeman circular polarisation, measured in two masers, yielded a value of
%\$|B_{||}|$ $\sim$ 45 mG. The authors also assumed a toroidal configuration
%\(applying a 1/R law) to derive a stellar surface magnetic field of $\sim$ 2.5
%\G. \\

By determining the ratio $\beta$ between the thermal pressure 
$P_{\rm th}=n_{\rm H}kT$ and the magnetic pressure 
$P_{\rm B}= B^2/8\pi$ we can estimate
the lower limit on the magnetic field's strength for which the field would
become dominant. \citet{Bujarrabal2002} provided a detailed analysis of OH
231.8+4.2 and its physical conditions and following their work we calculate
that in the dense central part of the PPN where $T=35$\,K and 
$n_{\rm H}=3\times10^6$\,cm$^{-3}$, the magnetic pressure dominates for
$B\geqslant0.6$\,mG. This threshold indicates that, assuming \citet{Ferreira2012}'s result, 
the magnetic pressure
largely dominates the thermal pressure at this location in this PPN\footnote{Note: water masers may come from denser condensations. They commonly show stronger field strength perhaps related to higher local densities.}. Similarly, based on
the work by \citet{Contreras2004} in the dense core of CRL 618 
($T=55$\,K and $n_{\rm H}=7.5\times10^6$\,cm$^{-3}$), the magnetic
pressure dominates for $B\geqslant$\,1.2\,mG, a value much lower than that
found by \citet{Herpin2009}. 
Although we estimated upper limits on the magnetic field for the molecular (neutral) areas in both PPNe (with low temperatures and high densities), the ionised regions (with higher temperatures) should not be discarded as the field still needs a partial ionisation to connect to the gas. Indeed in the outflows (showing high kinetic energy), the magnetic field might not govern the gas flow. We are therefore likely to see a variation of the flow-field relation inside the PPN.

\section{Conclusions}

 In this paper we present high resolution SMA submillimeter polarimetric observations of the proto-planetary nebulae CRL 618 and OH 231.8+4.2. In both cases we detected linear polarisation above 
$\sim3\sigma$ and a higher percentage polarisation is found in CRL 618 (C-rich) compared to OH 231.8+4.2 (O-rich). The difference is likely to be linked to the chemical nature (i.e. dust grain type) of the PPNe as also reported by \citet*{Sabin2007} for other evolved objects.

Also, the polarisation vectors indicate that the resulting magnetic field geometry supports the presence of a clear and well organised poloidal magnetic structure (i.e. aligned with the pairs of ionised outflows) in both PPNe. But while this configuration is more `simple' in CRL 618; the field shows an X-shaped structure in OH 231.8+4.2 and appears particularly patchy (although the field is organised in each individual component). However we do not discard the possibility that some vectors might belong to a toroidal field.

The observation of polarisation in the molecular lines, and particularly the CO lines, led to no positive detections above 3$\sigma$ in any of the PPN but we clearly observed, in both cases, an alignment of the ionised and molecular outflows with the magnetic field vectors, at large and small scales respectively, which supports the idea of a dynamically important field at small scale and of a probable magnetic launching origin for these outflows.

Finally we present a tentative explanation for the field's configuration evolution in late type stellar objects. 

The SMA probed to be an excellent tool to study the polarisation and magnetic field distribution in evolved stars. A better understanding of the role and evolution of magnetic fields in the late stages of stellar evolution will require more of these submillimeter and millimeter observations to investigate any evolutionary scheme or magnetic launching mechanism. In the future with the more sensitive polarisation capabilities of ALMA, we will be able to target a larger range of evolved objects particularly the faint PPNe and PNe.  

%But this type of investigation also require data on the the magnetic fields' strength to evaluate their importance from a dynamical point of view. This information would be brought by the polarisation of SiO masers for example\\

\section*{Acknowledgements}
The authors thank the SMA personal for supporting the observations as well as the referee for his/her comments and
suggestions, which significantly contributed to improving the quality of the publication. LS also thank S. Kemp for going through the paper and raising some issues. LS is supported by the CONACYT grant CB-2011-01-0168078 and RV is supported by PAPIIT-DGAPA-UNAM grant IN107914. This research has also made use of the SIMBAD database, operated at CDS, Strasbourg, France and the NASA's Astrophysics Data System. We acknowledge the use of observations made with the NASA/ESA Hubble Space Telescope, obtained from the data archive at the Space Telescope Science Institute. STScI is operated by the Association of Universities for Research in Astronomy, Inc. under NASA contract NAS 5-26555.

\bibliographystyle{mn2e}

\bibliography{sabin_SMA}

\begin{thebibliography}{}

\bibitem[\protect\citeauthoryear{{Alcolea}, {Bujarrabal}, {S{\'a}nchez
  Contreras}, {Neri} \& {Zweigle}}{{Alcolea} et~al.}{2001}]{Alcolea2001}
{Alcolea} J.,  {Bujarrabal} V.,  {S{\'a}nchez Contreras} C.,  {Neri} R.,
  {Zweigle} J.,  2001, \aap, 373, 932

\bibitem[\protect\citeauthoryear{{Balick}, {Huarte-Espinosa}, {Frank}, {Gomez},
  {Alcolea}, {Corradi} \& {Vinkovi{\'c}}}{{Balick} et~al.}{2013}]{Balick2013}
{Balick} B.,  {Huarte-Espinosa} M.,  {Frank} A.,  {Gomez} T.,  {Alcolea} J.,
  {Corradi} R.~L.~M.,    {Vinkovi{\'c}} D.,  2013, \apj, 772, 20

\bibitem[\protect\citeauthoryear{{Blackman}, {Frank}, {Markiel}, {Thomas} \&
  {Van Horn}}{{Blackman} et~al.}{2001}]{Blackman2001}
{Blackman} E.~G.,  {Frank} A.,  {Markiel} J.~A.,  {Thomas} J.~H.,    {Van Horn}
  H.~M.,  2001, \nat, 409, 485

\bibitem[\protect\citeauthoryear{{Bujarrabal}, {Alcolea}, {S{\'a}nchez
  Contreras} \& {Sahai}}{{Bujarrabal} et~al.}{2002}]{Bujarrabal2002}
{Bujarrabal} V.,  {Alcolea} J.,  {S{\'a}nchez Contreras} C.,    {Sahai} R.,
  2002, \aap, 389, 271

\bibitem[\protect\citeauthoryear{{Bujarrabal}, {Alcolea}, {Soria-Ruiz},
  {Planesas}, {Teyssier}, {Cernicharo}, {Decin}, {Dominik} \& et
  al.}{{Bujarrabal} et~al.}{2012}]{Bujarrabal2012}
{Bujarrabal} V.,  {Alcolea} J.,  {Soria-Ruiz} R.,  {Planesas} P.,  {Teyssier}
  D.,  {Cernicharo} J.,  {Decin} L.,  {Dominik} C.,    et al. 2012, \aap, 537,
  A8

\bibitem[\protect\citeauthoryear{{Bujarrabal}, {Alcolea}, {Soria-Ruiz},
  {Planesas}, {Teyssier}, {Marston}, {Cernicharo}, {Decin}, {Dominik},
  {Justtanont}, {de Koter}, {Melnick} \& et al.}{{Bujarrabal}
  et~al.}{2010}]{Bujarrabal2010}
{Bujarrabal} V.,  {Alcolea} J.,  {Soria-Ruiz} R.,  {Planesas} P.,  {Teyssier}
  D.,  {Marston} A.~P.,  {Cernicharo} J.,  {Decin} L.,  {Dominik} C.,
  {Justtanont} K.,  {de Koter} A.,  {Melnick} G.,    et al. 2010, \aap, 521, L3

\bibitem[\protect\citeauthoryear{{Chandrasekhar} \& {Fermi}}{{Chandrasekhar} \&
  {Fermi}}{1953}]{Chandra1953}
{Chandrasekhar} S.,  {Fermi} E.,  1953, \apj, 118, 113

\bibitem[\protect\citeauthoryear{{Chen}, {Rao}, {Wilner} \& {Liu}}{{Chen}
  et~al.}{2012}]{Chen2012}
{Chen} H.-R.,  {Rao} R.,  {Wilner} D.~J.,    {Liu} S.-Y.,  2012, \apjl, 751,
  L13

\bibitem[\protect\citeauthoryear{{Cho} \& {Lazarian}}{{Cho} \&
  {Lazarian}}{2005}]{Cho2005}
{Cho} J.,  {Lazarian} A.,  2005, \apj, 631, 361

\bibitem[\protect\citeauthoryear{{Choi}, {Brunthaler}, {Menten} \&
  {Reid}}{{Choi} et~al.}{2012}]{Choi2012}
{Choi} Y.~K.,  {Brunthaler} A.,  {Menten} K.~M.,    {Reid} M.~J.,  2012, in IAU
  Symposium Vol.~283 of IAU Symposium, {Trigonometric parallax of the
  protoplanetary nebula OH 231.8+4.2}.
pp 330--331

\bibitem[\protect\citeauthoryear{{Douchin}, {De Marco}, {Jacoby}, {Hillwig},
  {Frew}, {Bojicic}, {Jasniewicz} \& {Parker}}{{Douchin}
  et~al.}{2013}]{Douchin2013}
{Douchin} D.,  {De Marco} O.,  {Jacoby} G.~H.,  {Hillwig} T.~C.,  {Frew} D.~J.,
   {Bojicic} I.,  {Jasniewicz} G.,    {Parker} Q.~A.,  2013, in {Krzesi{\'n}}
  {ski} J.,  {Stachowski} G.,  {Moskalik} P.,   {Bajan} K.,  eds, 18th European
  White Dwarf Workshop. Vol.~469 of Astronomical Society of the Pacific
  Conference Series, {Testing the Binary Hypothesis for the Formation and
  Shaping of Planetary Nebulae}.
p.~293

\bibitem[\protect\citeauthoryear{{Etoka}, {Zijlstra}, {Richards}, {Matsuura} \&
  {Lagadec}}{{Etoka} et~al.}{2009}]{Etoka2009}
{Etoka} S.,  {Zijlstra} A.,  {Richards} A.~M.,  {Matsuura} M.,    {Lagadec} E.,
   2009, in {Murphy} S.~J.,  {Bessell} M.~S.,  eds, The Eighth Pacific Rim
  Conference on Stellar Astrophysics: A Tribute to Kam-Ching Leung Vol.~404 of
  Astronomical Society of the Pacific Conference Series, {The Geometrical and
  Magnetic Structure of the Proto-Planetary Nebula OH 231.8+4.2 Traced by OH
  Maser Emission}.
p.~311

\bibitem[\protect\citeauthoryear{{Falceta-Gon{\c c}alves}, {Lazarian} \&
  {Kowal}}{{Falceta-Gon{\c c}alves} et~al.}{2008}]{Falceta2008}
{Falceta-Gon{\c c}alves} D.,  {Lazarian} A.,    {Kowal} G.,  2008, \apj, 679,
  537

\bibitem[\protect\citeauthoryear{{Gardiner} \& {Frank}}{{Gardiner} \&
  {Frank}}{2001}]{Gardiner2001}
{Gardiner} T.~A.,  {Frank} A.,  2001, \apj, 557, 250

\bibitem[\protect\citeauthoryear{{Girart}, {Patel}, {Vlemmings} \&
  {Rao}}{{Girart} et~al.}{2012}]{Girart2012}
{Girart} J.~M.,  {Patel} N.,  {Vlemmings} W.~H.~T.,    {Rao} R.,  2012, \apjl,
  751, L20

\bibitem[\protect\citeauthoryear{{Goldreich} \& {Kylafis}}{{Goldreich} \&
  {Kylafis}}{1981}]{Goldreich1981}
{Goldreich} P.,  {Kylafis} N.~D.,  1981, \apjl, 243, L75

\bibitem[\protect\citeauthoryear{{Goldreich} \& {Kylafis}}{{Goldreich} \&
  {Kylafis}}{1982}]{Goldreich1982}
{Goldreich} P.,  {Kylafis} N.~D.,  1982, \apj, 253, 606

\bibitem[\protect\citeauthoryear{{Goodrich}}{{Goodrich}}{1991}]{Goodrich1991}
{Goodrich} R.~W.,  1991, \apj, 376, 654

\bibitem[\protect\citeauthoryear{{Greaves}}{{Greaves}}{2002}]{Greaves2002}
{Greaves} J.~S.,  2002, \aap, 392, L1

\bibitem[\protect\citeauthoryear{{Herpin}, {Baudy}, {Josselin}, {Thum} \&
  {Wiesemeyer}}{{Herpin} et~al.}{2009}]{Herpin2009}
{Herpin} F.,  {Baudy} A.,  {Josselin} E.,  {Thum} C.,    {Wiesemeyer} H.,
  2009, in {Strassmeier} K.~G.,  {Kosovichev} A.~G.,   {Beckman} J.~E.,  eds,
  IAU Symposium Vol.~259 of IAU Symposium, {Magnetic fields in AGB stars and
  (proto-) Planetary Nebulae}.
pp 47--52

\bibitem[\protect\citeauthoryear{{Hildebrand}}{{Hildebrand}}{1996}]{Hildebrand%
1996}
{Hildebrand} R.~H.,  1996, in {Roberge} W.~G.,  {Whittet} D.~C.~B.,  eds,
  Polarimetry of the Interstellar Medium Vol.~97 of Astronomical Society of the
  Pacific Conference Series, {Problems in Far-Infrared Polarimetry}.
p.~254

\bibitem[\protect\citeauthoryear{{Hildebrand}, {Dragovan} \&
  {Novak}}{{Hildebrand} et~al.}{1984}]{Hildebrand1984}
{Hildebrand} R.~H.,  {Dragovan} M.,    {Novak} G.,  1984, \apjl, 284, L51

\bibitem[\protect\citeauthoryear{{Ho}, {Moran} \& {Lo}}{{Ho}
  et~al.}{2004}]{Ho2004}
{Ho} P.~T.~P.,  {Moran} J.~M.,    {Lo} K.~Y.,  2004, \apjl, 616, L1

\bibitem[\protect\citeauthoryear{{Jones} \& {Spitzer} Jr.}{{Jones} \&
  {Spitzer}}{1967}]{Jones1967}
{Jones} R.~V.,  {Spitzer} Jr. L.,  1967, \apj, 147, 943

\bibitem[\protect\citeauthoryear{{Kastner}, {Weintraub}, {Zuckerman},
  {Becklin}, {McLean} \& {Gatley}}{{Kastner} et~al.}{1992}]{Kastner1992}
{Kastner} J.~H.,  {Weintraub} D.~A.,  {Zuckerman} B.,  {Becklin} E.~E.,
  {McLean} I.,    {Gatley} I.,  1992, \apj, 398, 552

\bibitem[\protect\citeauthoryear{{Kim} \& {Martin}}{{Kim} \&
  {Martin}}{1993}]{Kim1993}
{Kim} S.-H.,  {Martin} P.~G.,  1993, in American Astronomical Society Meeting
  Abstracts Vol.~25 of Bulletin of the American Astronomical Society, {The Size
  Distribution of Interstellar Dust Particles as Determined from Polarization}.
p.~1312

\bibitem[\protect\citeauthoryear{{Kim} \& {Martin}}{{Kim} \&
  {Martin}}{1995}]{Kim1995}
{Kim} S.-H.,  {Martin} P.~G.,  1995, \apj, 444, 293

\bibitem[\protect\citeauthoryear{{Koch}, {Tang} \& {Ho}}{{Koch}
  et~al.}{2012}]{Koch2012}
{Koch} P.~M.,  {Tang} Y.-W.,    {Ho} P.~T.~P.,  2012, \apj, 747, 79

\bibitem[\protect\citeauthoryear{{Kwok}}{{Kwok}}{1993}]{Kwok1993}
{Kwok} S.,  1993, \araa, 31, 63

\bibitem[\protect\citeauthoryear{{Kwok} \& {Bignell}}{{Kwok} \&
  {Bignell}}{1984}]{Kwok1984}
{Kwok} S.,  {Bignell} R.~C.,  1984, \apj, 276, 544

\bibitem[\protect\citeauthoryear{{Lazarian}}{{Lazarian}}{2003}]{Lazarian2003}
{Lazarian} A.,  2003, \jqsrt, 79, 881

\bibitem[\protect\citeauthoryear{{Lazarian} \& {Draine}}{{Lazarian} \&
  {Draine}}{2000}]{Lazarian2000}
{Lazarian} A.,  {Draine} B.~T.,  2000, \apjl, 536, L15

\bibitem[\protect\citeauthoryear{{Lazarian}, {Goodman} \& {Myers}}{{Lazarian}
  et~al.}{1997}]{Lazarian1997}
{Lazarian} A.,  {Goodman} A.~A.,    {Myers} P.~C.,  1997, \apj, 490, 273

\bibitem[\protect\citeauthoryear{{Lazarian} \& {Hoang}}{{Lazarian} \&
  {Hoang}}{2011}]{Lazarian2011}
{Lazarian} A.,  {Hoang} T.,  2011, in {Bastien} P.,  {Manset} N.,  {Clemens}
  D.~P.,   {St-Louis} N.,  eds, Astronomical Society of the Pacific Conference
  Series Vol.~449 of Astronomical Society of the Pacific Conference Series,
  {Alignment of Dust by Radiative Torque: Recent Developments}.
p.~116

\bibitem[\protect\citeauthoryear{{Leal-Ferreira}, {Vlemmings}, {Diamond},
  {Kemball}, {Amiri} \& {Desmurs}}{{Leal-Ferreira} et~al.}{2012}]{Ferreira2012}
{Leal-Ferreira} M.~L.,  {Vlemmings} W.~H.~T.,  {Diamond} P.~J.,  {Kemball} A.,
  {Amiri} N.,    {Desmurs} J.-F.,  2012, \aap, 540, A42

\bibitem[\protect\citeauthoryear{{Leal-Ferreira}, {Vlemmings}, {Kemball} \&
  {Amiri}}{{Leal-Ferreira} et~al.}{2013}]{LealFerreira2013}
{Leal-Ferreira} M.~L.,  {Vlemmings} W.~H.~T.,  {Kemball} A.,    {Amiri} N.,
  2013, ArXiv e-prints

\bibitem[\protect\citeauthoryear{{Lee}, {Sahai}, {S{\'a}nchez Contreras},
  {Huang} \& {Hao Tay}}{{Lee} et~al.}{2013}]{Lee2013b}
{Lee} C.-F.,  {Sahai} R.,  {S{\'a}nchez Contreras} C.,  {Huang} P.-S.,    {Hao
  Tay} J.~J.,  2013, ArXiv e-prints

\bibitem[\protect\citeauthoryear{{Lee}, {Yang}, {Sahai} \& {S{\'a}nchez
  Contreras}}{{Lee} et~al.}{2013}]{Lee2013a}
{Lee} C.-F.,  {Yang} C.-H.,  {Sahai} R.,    {S{\'a}nchez Contreras} C.,  2013,
  \apj, 770, 153

\bibitem[\protect\citeauthoryear{{Lequeux} \& {Jourdain de Muizon}}{{Lequeux}
  \& {Jourdain de Muizon}}{1990}]{Lequeux1990}
{Lequeux} J.,  {Jourdain de Muizon} M.,  1990, \aap, 240, L19

\bibitem[\protect\citeauthoryear{{Maldoni}, {Egan}, {Robinson}, {Smith} \&
  {Wright}}{{Maldoni} et~al.}{2004}]{Maldoni2004}
{Maldoni} M.~M.,  {Egan} M.~P.,  {Robinson} G.,  {Smith} R.~G.,    {Wright}
  C.~M.,  2004, \mnras, 349, 665

\bibitem[\protect\citeauthoryear{{Marrone} \& {Rao}}{{Marrone} \&
  {Rao}}{2008}]{Marrone2008}
{Marrone} D.~P.,  {Rao} R.,  2008, in Society of Photo-Optical Instrumentation
  Engineers (SPIE) Conference Series Vol.~7020 of Society of Photo-Optical
  Instrumentation Engineers (SPIE) Conference Series, {The submillimeter array
  polarimeter}

\bibitem[\protect\citeauthoryear{{Martin-Pintado}, {Gaume}, {Bachiller} \&
  {Johnson}}{{Martin-Pintado} et~al.}{1993}]{Pintado1993}
{Martin-Pintado} J.,  {Gaume} R.,  {Bachiller} R.,    {Johnson} K.,  1993,
  \apj, 419, 725

\bibitem[\protect\citeauthoryear{{Mathis}}{{Mathis}}{1986}]{Mathis1986}
{Mathis} J.~S.,  1986, \apj, 308, 281

\bibitem[\protect\citeauthoryear{{Matt}, {Balick}, {Winglee} \&
  {Goodson}}{{Matt} et~al.}{2000}]{Matt2000}
{Matt} S.,  {Balick} B.,  {Winglee} R.,    {Goodson} A.,  2000, \apj, 545, 965

\bibitem[\protect\citeauthoryear{{Matthews}, {Fiege} \&
  {Moriarty-Schieven}}{{Matthews} et~al.}{2002}]{Matthews2002}
{Matthews} B.~C.,  {Fiege} J.~D.,    {Moriarty-Schieven} G.,  2002, \apj, 569,
  304

\bibitem[\protect\citeauthoryear{{Matthews}, {Wilson} \& {Fiege}}{{Matthews}
  et~al.}{2001}]{Matthews2001}
{Matthews} B.~C.,  {Wilson} C.~D.,    {Fiege} J.~D.,  2001, \apj, 562, 400

\bibitem[\protect\citeauthoryear{{Morgan}}{{Morgan}}{1995}]{Morgan1995}
{Morgan} J.~A.,  1995, in {Shaw} R.~A.,  {Payne} H.~E.,   {Hayes} J.~J.~E.,
  eds, Astronomical Data Analysis Software and Systems IV Vol.~77 of
  Astronomical Society of the Pacific Conference Series, {WIP -- an Interactive
  Graphics Software Package}.
p.~129

\bibitem[\protect\citeauthoryear{{Morris}, {Guilloteau}, {Lucas} \&
  {Omont}}{{Morris} et~al.}{1987}]{Morris1987}
{Morris} M.,  {Guilloteau} S.,  {Lucas} R.,    {Omont} A.,  1987, \apj, 321,
  888

\bibitem[\protect\citeauthoryear{{Morris}, {Lucas} \& {Omont}}{{Morris}
  et~al.}{1985}]{Morris1985}
{Morris} M.,  {Lucas} R.,    {Omont} A.,  1985, \aap, 142, 107

\bibitem[\protect\citeauthoryear{{Nakashima}, {Fong}, {Hasegawa}, {Hirano},
  {Koning}, {Kwok}, {Lim}, {Dinh-Van-Trung} \& {Young}}{{Nakashima}
  et~al.}{2007}]{Nakashima2007}
{Nakashima} J.-i.,  {Fong} D.,  {Hasegawa} T.,  {Hirano} N.,  {Koning} N.,
  {Kwok} S.,  {Lim} J.,  {Dinh-Van-Trung}   {Young} K.,  2007, \aj, 134, 2035

\bibitem[\protect\citeauthoryear{{Nordhaus}}{{Nordhaus}}{2008}]{Nordhaus2008}
{Nordhaus} J.~T.,  2008, PhD thesis, University of Rochester

\bibitem[\protect\citeauthoryear{{Ostriker}, {Stone} \& {Gammie}}{{Ostriker}
  et~al.}{2001}]{Ostriker2001}
{Ostriker} E.~C.,  {Stone} J.~M.,    {Gammie} C.~F.,  2001, \apj, 546, 980

\bibitem[\protect\citeauthoryear{{Padovani}, {Brinch}, {Girart},
  {J{\o}rgensen}, {Frau}, {Hennebelle}, {Kuiper}, {Vlemmings}, {Bertoldi},
  {Hogerheijde}, {Juhasz} \& {Schaaf}}{{Padovani} et~al.}{2012}]{Padovani2012}
{Padovani} M.,  {Brinch} C.,  {Girart} J.~M.,  {J{\o}rgensen} J.~K.,  {Frau}
  P.,  {Hennebelle} P.,  {Kuiper} R.,  {Vlemmings} W.~H.~T.,  {Bertoldi} F.,
  {Hogerheijde} M.,  {Juhasz} A.,    {Schaaf} R.,  2012, \aap, 543, A16

\bibitem[\protect\citeauthoryear{{P{\'e}rez-S{\'a}nchez}, {Vlemmings}, {Tafoya}
  \& {Chapman}}{{P{\'e}rez-S{\'a}nchez} et~al.}{2013}]{Perez2013}
{P{\'e}rez-S{\'a}nchez} A.~F.,  {Vlemmings} W.~H.~T.,  {Tafoya} D.,
  {Chapman} J.~M.,  2013, ArXiv e-prints

\bibitem[\protect\citeauthoryear{{Phillips}, {Williams}, {Mampaso} \&
  {Ukita}}{{Phillips} et~al.}{1992}]{Phillips1992}
{Phillips} J.~P.,  {Williams} P.~G.,  {Mampaso} A.,    {Ukita} N.,  1992, \aap,
  260, 283

\bibitem[\protect\citeauthoryear{{Rao} \& {Marrone}}{{Rao} \&
  {Marrone}}{2005}]{Rao2005}
{Rao} R.,  {Marrone} D.~P.,  2005, in {Adamson} A.,  {Aspin} C.,  {Davis} C.,
  {Fujiyoshi} T.,  eds, Astronomical Polarimetry: Current Status and Future
  Directions Vol.~343 of Astronomical Society of the Pacific Conference Series,
  {Polarization with the Submillimeter Array (SMA)}.
p.~59

\bibitem[\protect\citeauthoryear{{Sabin}, {Zijlstra} \& {Greaves}}{{Sabin}
  et~al.}{2007}]{Sabin2007}
{Sabin} L.,  {Zijlstra} A.~A.,    {Greaves} J.~S.,  2007, \mnras, 376, 378

\bibitem[\protect\citeauthoryear{{Sahai}, {Morris}, {S{\'a}nchez Contreras} \&
  {Claussen}}{{Sahai} et~al.}{2007}]{Sahai2007}
{Sahai} R.,  {Morris} M.,  {S{\'a}nchez Contreras} C.,    {Claussen} M.,  2007,
  \aj, 134, 2200

\bibitem[\protect\citeauthoryear{{S{\'a}nchez Contreras}, {Alcolea},
  {Bujarrabal} \& {Neri}}{{S{\'a}nchez Contreras} et~al.}{1998}]{Sanchez1998}
{S{\'a}nchez Contreras} C.,  {Alcolea} J.,  {Bujarrabal} V.,    {Neri} R.,
  1998, \aap, 337, 233

\bibitem[\protect\citeauthoryear{{S{\'a}nchez Contreras}, {Bujarrabal} \&
  {Alcolea}}{{S{\'a}nchez Contreras} et~al.}{1997}]{Sanchez1997}
{S{\'a}nchez Contreras} C.,  {Bujarrabal} V.,    {Alcolea} J.,  1997, \aap,
  327, 689

\bibitem[\protect\citeauthoryear{{S{\'a}nchez Contreras}, {Bujarrabal},
  {Castro-Carrizo}, {Alcolea} \& {Sargent}}{{S{\'a}nchez Contreras}
  et~al.}{2004}]{Contreras2004}
{S{\'a}nchez Contreras} C.,  {Bujarrabal} V.,  {Castro-Carrizo} A.,  {Alcolea}
  J.,    {Sargent} A.,  2004, \apj, 617, 1142

\bibitem[\protect\citeauthoryear{{S{\'a}nchez Contreras}, {Gil de Paz} \&
  {Sahai}}{{S{\'a}nchez Contreras} et~al.}{2004}]{Sanchez2004b}
{S{\'a}nchez Contreras} C.,  {Gil de Paz} A.,    {Sahai} R.,  2004, \apj, 616,
  519

\bibitem[\protect\citeauthoryear{{S{\'a}nchez Contreras} \&
  {Sahai}}{{S{\'a}nchez Contreras} \& {Sahai}}{2004}]{Sanchez2004}
{S{\'a}nchez Contreras} C.,  {Sahai} R.,  2004, \apj, 602, 960

\bibitem[\protect\citeauthoryear{{S{\'a}nchez Contreras}, {Sahai} \& {Gil de
  Paz}}{{S{\'a}nchez Contreras} et~al.}{2002}]{Sanchez2002}
{S{\'a}nchez Contreras} C.,  {Sahai} R.,    {Gil de Paz} A.,  2002, \apj, 578,
  269

\bibitem[\protect\citeauthoryear{{Sault}, {Teuben} \& {Wright}}{{Sault}
  et~al.}{2011}]{Sault2011}
{Sault} R.~J.,  {Teuben} P.~J.,    {Wright} M.~C.~H.,  2011, Astrophysics
  Source Code Library, p.~6007

\bibitem[\protect\citeauthoryear{{Soker}}{{Soker}}{2004}]{Soker2004}
{Soker} N.,  2004, in {Meixner} M.,  {Kastner} J.~H.,  {Balick} B.,   {Soker}
  N.,  eds, Asymmetrical Planetary Nebulae III: Winds, Structure and the
  Thunderbird Vol.~313 of Astronomical Society of the Pacific Conference
  Series, {Shaping Planetary Nebulae and Related Objects}.
p.~562

\bibitem[\protect\citeauthoryear{{Tafoya}, {Loinard}, {Fonfr{\'{\i}}a},
  {Vlemmings}, {Mart{\'{\i}}-Vidal} \& {Pech}}{{Tafoya}
  et~al.}{2013}]{Tafoya2013}
{Tafoya} D.,  {Loinard} L.,  {Fonfr{\'{\i}}a} J.~P.,  {Vlemmings} W.~H.~T.,
  {Mart{\'{\i}}-Vidal} I.,    {Pech} G.,  2013, ArXiv e-prints

\bibitem[\protect\citeauthoryear{{Tocknell}, {De Marco} \& {Wardle}}{{Tocknell}
  et~al.}{2013}]{Tocknell2013}
{Tocknell} J.,  {De Marco} O.,    {Wardle} M.,  2013, ArXiv e-prints

\bibitem[\protect\citeauthoryear{{Vlemmings}}{{Vlemmings}}{2011}]{Vlemmings201%
1}
{Vlemmings} W.~H.~T.,  2011, in Asymmetric Planetary Nebulae 5 Conference
  {Magnetic fields around (post-)AGB stars and (Pre-)Planetary Nebulae}

\bibitem[\protect\citeauthoryear{{Vlemmings}, {Ramstedt}, {Rao} \&
  {Maercker}}{{Vlemmings} et~al.}{2012}]{Vlemmings2012}
{Vlemmings} W.~H.~T.,  {Ramstedt} S.,  {Rao} R.,    {Maercker} M.,  2012, \aap,
  540, L3

\bibitem[\protect\citeauthoryear{{Woods}, {Nyman}, {Sch{\"o}ier}, {Zijlstra},
  {Millar} \& {Olofsson}}{{Woods} et~al.}{2005}]{Woods2005}
{Woods} P.~M.,  {Nyman} L.-{\AA}.,  {Sch{\"o}ier} F.~L.,  {Zijlstra} A.~A.,
  {Millar} T.~J.,    {Olofsson} H.,  2005, \aap, 429, 977

\bibitem[\protect\citeauthoryear{{Wright} \& {Sault}}{{Wright} \&
  {Sault}}{1993}]{Wright1993}
{Wright} M.~C.~H.,  {Sault} R.~J.,  1993, \apj, 402, 546

\end{thebibliography}
\vspace{-0.3cm}

%and the anonymous referee for the comments that significantly improved the paper
\bsp
\label{lastpage}

\end{document}